\documentclass[11pt,sort&compress,fleqn]{cas-sc}

\usepackage{amssymb}

\usepackage{url}
\usepackage{xcolor}

\usepackage{amsthm}
\usepackage{amsfonts}
\usepackage{amsmath} 
\usepackage{verbatim} 	       
\usepackage{setspace}
\usepackage{mdframed}
\usepackage{enumitem}		   
\usepackage{subcaption}		   
\usepackage{hyperref}
\usepackage{placeins}		   
\usepackage{siunitx} 		   
\usepackage{numprint}		   
\usepackage{hhline}			   
\usepackage[ruled,vlined]{algorithm2e} 	   

\usepackage{natbib}
\def\tsc#1{\csdef{#1}{\textsc{\lowercase{#1}}\xspace}}
\tsc{WGM}
\tsc{QE}

\usepackage[demo]{adjustbox}
\usepackage{xcolor}
\usepackage{atbegshi}
\AtBeginDocument{\AtBeginShipoutNext{\AtBeginShipoutDiscard}}	
\PassOptionsToPackage{hyphens}{url}\usepackage{hyperref} 

\begin{document}
	
	\pagestyle{empty} 
	\begin{titlepage}
		\color[rgb]{.4,.4,1}
		\hspace{5mm}

		\bigskip
		
		\hspace{15mm}
		\begin{minipage}{10mm}
			\color[rgb]{.7,.7,1}
			\rule{1pt}{226mm}
		\end{minipage}
		\begin{minipage}{133mm}
			\vspace{10mm}        
			\color{black}
			\sffamily
			\LARGE\bfseries Carbon Nanotubes as a Basis of  \\[-0.3\baselineskip] Metamaterials and Nanostructures:  \\[-0.3\baselineskip] Crafting via Design Optimization\\[-0.3\baselineskip] 
			
			\vspace{5mm}
			{\large {Preprint of the article published in \\[-0.4\baselineskip] Mechanics of Materials (2024) }} 
			
			\vspace{10mm}        
			{\large Marko \v{C}ana\dj{}ija, Stefan Ivi\'{c} } 
			
			\large
			
			\vspace{40mm}
			\vspace{5mm}
			
			\small
			\url{https://doi.org/10.1016/j.mechmat.2024.105105}
			
			\textcircled{c} 2024. This manuscript version is made available under the CC-BY-NC-ND 4.0 license \url{http://creativecommons.org/licenses/by-nc-nd/4.0/}
			\hspace{30mm} 
			\color[rgb]{.4,.4,1} 
			\includegraphics[width=3cm]{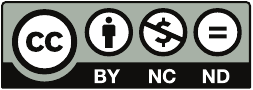}        
		\end{minipage}
	\end{titlepage}

\title [mode = title]{Carbon Nanotubes as a Basis of Metamaterials and Nanostructures: Crafting via Design Optimization}  
\shorttitle{Carbon Nanotubes as a Basis of Metamaterials and Nanostructures}    
\shortauthors{M. \v{C}ana\dj{}ija et al.}

\author[1]{Marko \v{C}ana\dj{}ija}[
       auid=000,
       orcid=0000-0001-6550-0258]
\cormark[1]
\ead{marko.canadija@riteh.uniri.hr}
\credit{Conceptualization, Funding acquisition, Investigation, Methodology, Project administration, Software, Validation, Writing - Original Draft, Writing - Review \& Editing.}
\affiliation[1]{organization={University of Rijeka, Faculty of Engineering},
	addressline={Vukovarska 58}, 
	city={Rijeka},
	postcode={51000}, 
	country={Croatia}}
\author[1]{Stefan Ivi\'{c}}
\credit{Investigation, Methodology, Software, Validation, Writing - Original Draft, Writing - Review \& Editing}
\cortext[1]{Corresponding author}

\begin{abstract}
Nanotruss structures made of carbon nanotubes are investigated in two conceptual applications: either as building blocks of metamaterials or for nanostructural applications. The nanotrusses are optimized for different purposes, including various loadings, boundary conditions, parameterizations, objectives and constraints used to formulate optimization problems. The procedure relies on a recently developed framework consisting of molecular dynamics simulations, neural networks and finite elements. This framework is now used in the design optimization of nanostructures and the performances of different popular heuristic optimization methods are compared. Five applications of nanotrusses made of carbon nanotubes are analyzed in detail to investigate the mechanical behavior of such structures and the efficiency of the optimizations. Besides an introductory example, the design of an energy trapping carbon nanotube nanotruss, an auxetic nanotruss, a cantilever nanotruss and the maximization of the compressive strength of a metamaterial are presented. It is shown that the exceptional mechanical properties of carbon nanotubes can indeed be exploited for the development of structures and materials with extraordinary mechanical properties. Although hampered by material and geometrical nonlinearity of the problem, most of the tested optimization methods have proven to be a good choice for the design of such materials and structures.
\end{abstract}
 


\maketitle

\section{Introduction}
\label{sec_Introduction}
The development of new materials has always been a very active area of engineering. For the present research, the period after the Second World War is of interest, when a new class of materials emerged - metamaterials. These are a special class of materials designed to have a particular property that rarely occurs in nature. Originally the research was focused on microwave technology and shaping of antenna beams, but today this is a rapidly developing field that encompasses many technical areas. Although the focus of the present research is only on mechanical properties, other types of properties are often considered in the literature. These are usually interdisciplinary in nature and include thermal, electromagnetic, optical, acoustic and other properties that differ significantly from those of conventional materials.

While many different building blocks can be used as the basic structure of a metamaterial, carbon nanotubes (CNT) are a particularly promising candidate due to their exceptional material properties. Motivated by the nanotrusses fabricated of non-CNT-based materials \cite{meza2014mechanical,meza2015resilient} the present research envisages nanotrusses fabricated of CNT. Up to now, only various simplest cubic-based nanotrusses are investigated, often analyzing only the peculiarities of the interconnection of CNT in a nanotruss. Such nanotrusses are almost exclusively analyzed using molecular dynamics, very rarely using finite elements (FE). For example, an MD study \cite{zhang2018nano} compares the mechanical and thermal properties of simple cubic (SC) and face-centered cubic (FCC) nanotruss architectures consisting of (8,8) nanotubes. It was found that very short junctions connecting the CNT have a significant impact on the properties and that the FCC architecture shows superior performance. Alternatively, fullerene can be used instead of junctions to connect CNT \cite{wu2013fracture,pedrielli2017designing}.

Apart from connections, the studies often focus on the development of a specific property. In \cite{cai2021hierarchical}, a framework for tunable mechanical and thermal properties of kirigami metamaterials based on CNT and graphene is presented. The approach involves the application of MD and finite elements, but no optimization was performed. The same group analyses seven types of biomaterial structures as a basis for carbon nanoarchitected materials \cite{cai2022lessons}. The beam finite elements are matched with the MD results and used for CNT members. CNT nanotruss based metamaterials with negative Poisson's ratio were developed in \cite{cai2021hierarchical,wu2013fracture,pedrielli2017designing,cai2023tunable}, again using an intuitive design. At the end, it should be mentioned that the recent graphene-based origami metamaterials open up new possibilities for tailored properties, for example the elastocaloric effect \cite{Cai2023} and energy harvesting \cite{Yan2024}.

As far as experimental research is concerned, this is still limited to the simplest CNT nanotrusses. In \cite{portela2021supersonic}, it is experimentally shown that the nanoarchitected pyrolytic carbon metamaterial with tetrakaidecahedral structure can absorb up to 72 \% more impact energy than Kevlar composites. Polymer composites based on CNT trusses were manufactured and tested in  \cite{meaud2014simultaneously}. Experiments show that they simultaneously exhibit high stiffness and high damping properties. Carbon aerogels are not exactly truss structures, but they are closely related. They are very light and sparse structures that use either CNT or graphene to improve mechanical properties. For example, \cite{sun2013multifunctional} is developing an ultralight aerogel with a specific mass density of $\rho=0.16$ kg/m$^3$, which has a tensile strength of about $\sigma_\mathrm{max}=11$ kPa, giving a specific tensile strength of $\sigma_\mathrm{max}/\rho=0.06875$ MPa/kg, quite similar to that of conventional steel. 

The above brief overview of the state of the art shows that CNT nanotrusses are indeed very interesting elements for new metamaterials. However, all the  existing structures mentioned above are limited to very simple architectures, mainly of the cubic type. The main advantage of this simple architecture is perhaps that such structures are easier to fabricate and that they can be developed with an intuitive approach. On the other hand, the complex CNT nanotrusses offer a wider range of possibilities but suffer from a more complex fabrication and development. Due to such complex nature, the intuitive approach to design is no longer appropriate. One must rely on more complex procedures involving optimization techniques. 

There are a variety of options when it comes to choosing optimization techniques. More classical local search methods that resort to the evaluation of gradients and possibly Hessians can be computationally too expensive due to numerical approximations of these derivatives. This is especially true for optimization problems with a large number of variables and are also computationally demanding due to material and geometric nonlinearity. On the other hand, the nonlinearity and non-local effects make the objective and constraint functions non-smooth and complicated to solve using local search methods. In this sense, heuristic techniques may be a more advantageous choice. Since this research deals with truss-like structures, it is interesting to give a brief overview of truss optimization techniques.

When it comes to optimizing truss structures, size, shape and topology optimization are performed. All of these are quite popular research topics, and a detailed literature review will not be conducted here. Instead, only a brief overview is given. In size and shape optimization, the variables are the cross-sectional areas and the node positions, respectively. In topology optimization, new nodes and finite elements are introduced. The least demanding problems are usually size optimizations, and most optimization methods work well for majority of these problems. At the other end of the spectrum is topology optimization, which is a discontinuous problem and thus a hard task for any optimization method. Size and topology optimization will not be discussed in this paper; for a more detailed presentation of the concepts, the interested reader is referred to \cite{tejani2018size} for a start. 

This research is based on the shape optimization of trusses. This is often a continuous problem, e.g. minimizing the truss mass while keeping the stresses within certain limits by repositioning nodes. The repositioning can be done either by parameterizing the truss geometry or by considering the nodal offsets as the optimization variables. However, one point should be noted: most studies deal with material and geometric linear trusses, which is a significant simplification. The assumption of linearity severely limits possible CNT applications. Further, to the best of our knowledge, there are no attempts in the literature to optimize a larger nanotruss made of CNT, especially not with respect to material and geometric nonlinear behavior.

Although there have been attempts to tune mechanical properties of carbon nanotube-based metamaterials with basic cubic trusses (SC, FCC and the like) using intuition, an approach based on the nanotruss shape optimization has not been considered so far. The reason is that the optimization procedure requires a large number of repetitive structural simulations, so MD is not a viable approach due to its high computational cost. A popular finite element approach - molecular structural mechanics (MSM), as proposed in \cite{Li03}, which relies on the harmonic potential to model atomic bonds as beams, is quite simple, see \cite{Canadija17,papadopoulos2018neural}. In particular, the harmonic potential used in such models leads to a linear behavior and the adaptation to more realistic MD simulations can be a problem. This is especially true for problems at room temperature, as the stochastic thermal vibrations of the atoms are neglected. There are also many open questions in connection with another alternative - non-local analytical rod/beam elements, for an overview see \cite{canadija2024computational, Canadija2023}.

The best aspects of MD and MSM approaches are combined in a new open-access framework comprising MD, neural networks (NN) and FE recently developed in \cite{canadija2024computational} and \cite{canadija2021deep,kosmerl2022}. It is shown that FE, with the aid of NN, can replace MD simulations of CNT nanostructures while maintaining the required MD accuracy and that at the same time the calculations are very efficient and robust. This enables a large number of evaluations required for the chosen optimization scheme. 

The MD-NN-FE model mentioned above represents a computational tool that is used as a starting point for the present study. Here it becomes part of an optimization procedure so that, for the first time, complex nanotrusses made of carbon nanotubes can be optimally designed. To illustrate the possibilities of the new framework, its capabilities are used to develop three metamaterials with tailored properties. These are an energy trapping application, an auxetic material and maximization of the compressive strength of a metamaterial. Most of the underlying design concepts mentioned above come from the macroscale counterparts, however, relying on the truss-like structures instead of macroscopic solids or beams to achieve such behavior is a new approach. Besides searching for tailored metamaterials, the framework is also being tested for shape optimization of two nanotrusses that can be used in nanotechnological structural applications. Although the examples may seem unrelated at first glance, it is precisely this diversity that enables the generalizability of the method by rigorously testing multiple heuristic optimization methods to determine the one that is best suited to the problem at hand. This is not a trivial problem, as the problem is highly complex due to the inclusion of material nonlinearity in the form of nonlinear stress-strain tension-compression curves and geometric nonlinearity due to the large displacements involved.

\section{Methodology}
\label{sec_Methods}
\begin{figure}
	\centering
	\includegraphics[scale=0.07]{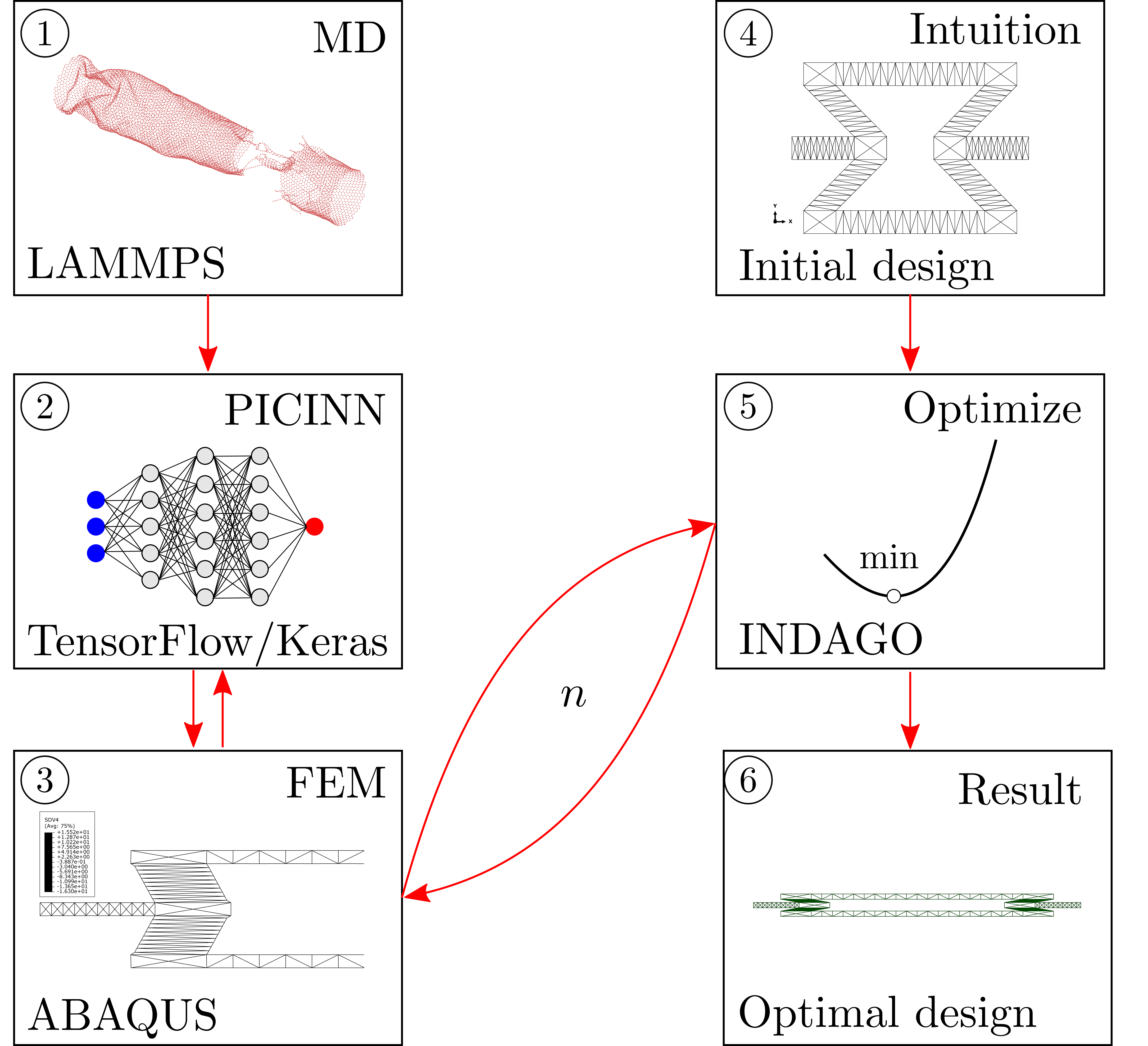}
	\caption{The optimization framework for CNT nanotrusses.} 
	\label{Fig_1_framework}
\end{figure}

The proposed optimization framework consists of several building blocks as shown in Fig.~\ref{Fig_1_framework}. The core is the tailored truss finite element that can reproduce the uniaxial mechanical behavior of single-walled carbon nanotubes (SWCNT) at the room temperature as introduced in a series of papers \cite{canadija2021deep,kosmerl2022,canadija2024computational}. Firstly, a comprehensive set of molecular dynamics simulations of uniaxial tensile \cite{canadija2021deep} and compressive \cite{canadija2024computational} tests at the room temperature were performed using LAMMPS \cite{Plimpton1995}. In the second step, the obtained stress-strain curves data was used as a dataset for training neural networks \cite{kosmerl2022}. It turned out that such neural networks can capture stress-strain curves with very high accuracy and incorporate the effects of stochastic vibrations due to temperature. This actually represents an obstacle to a successful finite element implementation. This is circumvented in \cite{canadija2024computational}, which includes a series of verification problems and confirms that the order of accuracy of the new nanotruss finite element is the same as that of molecular dynamics. 

The nanotruss finite elements at hand are now used with the optimization package Indago \cite{indago2024} to obtain the present framework. In short, a Python code that is based on the Indago package has been developed. At each evaluation during optimization, using selected design variables, an automatic generation of the input file for Abaqus is carried out. This includes the generation of the FE mesh, boundary conditions and calculation parameters. After this, the Abaqus analysis is invoked, and upon completion, the objective function and the values of the constraints are calculated in a post-processing step. All this data is returned to Indago, and then the next evaluation follows until the optimization is completed. 

The main aspects of the above framework are summarized in the following sections, while the detailed insight can be gained in the references given above.

\subsection{Molecular Dynamics}
\label{ssec_MD} 
To obtain a reliable description of the mechanical behavior of SWCNT in tension and compression, a series of molecular dynamics (MD) simulations was performed in LAMMPS. The series consisted of 818 different configurations with all possible CNT chirialities defining nanotubes with diameters up to 4.0 nm. This covers chiralities $(n,m)$ from (3, 3) to (51, 0). Each configuration was tested three times in tension and three times in compression, and the results were averaged. As explained in \cite{canadija2021deep}, each of these simulations starts with different random velocity fields. This is necessary to mitigate the stochastic nature of the problem due to thermal fluctuations. Therefore, 4908 MD simulations were performed. The length-to-diameter ratio was approximately 5 in all cases. In order to obtain practically usable results, all simulations were carried out at room temperature. On the downside, the temperature induces stochastic vibrations of the atoms, which add noise to the results and thus considerably increase the complexity of the problem.

The potential used is the modified AIREBO potential as described in \citep{Shenderova2000}. This is nowadays a classical model that is often considered very accurate \cite{qian2021comprehensive}. Note that the recently developed potentials are obtained by applying machine learning techniques, which further increases the conformity to \textit{ab initio} calculations \citep{canadija2021deep}. Unfortunately, these potentials are quite slow in practical implementation, which makes their application impractical in the present case.

As already mentioned, both tensile and compressive behavior are analyzed. In compression, carbon nanotubes show an almost linear stress-strain behavior, while this behavior is clearly nonlinear in tension  \cite{canadija2024computational}. The compressive strength is also greatly reduced compared to the tensile strength. This can be explained by the fact that carbon nanotubes are generally significantly affected by buckling. In this case, the buckling can be of a shell-like type, beam-like or rope-like \citep{wang2010recent}. Which buckling type is active depends on the ratio between the length $L$ and the diameter $D$. The limit point between the shell-like and the beam-like type is at $L/D=12.5$ \citep{buehler2004deformation}. Since in the present case MD tests were performed with $L/D\approx5$, all SWCNT fail due to shell-like buckling. The reader is reminded that the diameter plays a major role in such buckling, while the length is almost irrelevant. Therefore, the larger the diameter, the lower the buckling load. 

On the other hand, CNTs can withstand very high stresses under tension, often larger than 100 GPa. A comprehensive MD-NN study \cite{canadija2021deep} showed that not only the tensile strength, but also the Young's modulus, ultimate tensile strain and Possion's ratio depend mainly on the chirality, while the diameter is of secondary importance. Moreover, the complete lack of influence of the diameter on the ultimate tensile strain leads to the following conclusion. In contrast to the Young's modulus, tensile strength and Poisson's ratio, the only property for which the diameter is not directly included in the calculation is the ultimate tensile strain. Nevertheless, the diameter is taken into account indirectly via the chirality. This raises the question of whether the calculation of the diameter is the main cause of the diameter dependence of the other three properties that is present in SWCNTs with smaller diameters. In particular, the cross-section of a nanotube is an irregular polygon approximated as a thin ring. As the diameter increases, the quality of the approximation also increases and the dependence on the diameter disappears. Finally, it is emphasized that recent experimental research \cite{sun2024experimental} has come to similar conclusions about SWCNTs. For multi-walled carbon nanotubes, it remains to be seen how pronounced this effect is.  

All tests were performed by prescribing the velocity at one end of the SWCNT while the other end remained fixed. The applied velocities were adjusted so that the strain rate remained constant during all tests. The resulting true stress-true strain curves were obtained using a standard nanotube thickness of 0.34 nm. Finally, it should be emphasized once more that these curves as well as their endpoints depend mainly on the chirality and not on the diameter.

\subsection{Neural Networks}
\label{ssec_NN} 
After the MD simulations were performed, a dataset consisting of the chirality parameters $n$ and $m$, the true strain $\epsilon$ as inputs and the true stress $\sigma$ as output was obtained. As will be shown later, an additional dataset with the current diameter $D$ as output was also required, i.e. $\lbrace n,m,\epsilon,D\rbrace$. Each of the two datasets consisted of about 1.5 million points. As already mentioned, an initial attempt to model $\lbrace n,m,\epsilon,\sigma\rbrace$ showed that a classical feed forward network or a convolutional neural network can capture the stress-strain behavior with surprising accuracy \cite{kosmerl2022}, capturing even the smallest variations in stress-strain curves caused by the stochastic thermal vibrations of the atoms. Unfortunately, this proves to be a disadvantage in the subsequent implementation of the finite elements. In particular, the thermal vibrations of the atoms cause vibrations of the complete CNT, which in turn leads to small oscillations in the stress as the strain increases. This apparent loss of monotonic increase of the $\sigma$-$\epsilon$ curve has two effects: first, it causes the tangential modulus to oscillate and take on both negative and positive values. Since the problem at hand is highly nonlinear in both the material and geometric sense, the nonlinear finite element framework cannot converge due to such oscillations. Secondly, the underlying physical principle that the strain energy should be a convex function of strain is violated.

For this reason, a simple application of the neural network as a substitute for the constitutive model cannot be pursued. To restore the convexity of the strain energy, which will also solve the problem of the alternating tangential modulus, a somewhat upgraded approach to the one in \cite{huang2022variational} was proposed in \cite{canadija2024computational}. More specifically, partially input convex integrable neural networks (PICINN) were used to enforce the convexity of the strain energy $\psi$ with respect to the input variable $\epsilon$, while such constraints were not enforced with respect to the chiralities $n,m$. In addition, the physically justified requirement is imposed that the stress and strain energy vanish in the absence of strain, i.e. $\sigma=0$ and $\psi=0$ at $\epsilon=0$. All these constraints are met by choosing a specific architecture of the NN. Such a NN then outputs the strain energy, but by applying the automatic differentiation to obtain $\sigma=\partial_\epsilon \psi(\epsilon,n,m)$, the training is performed with the original dataset $\lbrace n,m,\epsilon,\sigma\rbrace$. In this way, the trained NN can then be used as a constitutive model for a variety of problems. The NN was developed using Keras and TensorFlow packages.

In addition, \cite{canadija2024computational} shows that NN captures the MD dataset with a high degree of accuracy in both compression and tension. Due to the automatic differentiation and integration, the proposed NN design enables the calculation of strain energy, stresses and tangent modulus by a single neural network. While the other details can be found in the given reference, perhaps one result can be pointed out. A unique feature of this approach is that the NN uses both the compression and tension data to determine the Young's modulus in the relaxed state of the carbon nanotube. This differs from most results published in the literature, which rely only on the tensile data. When done in this way, the Young's modulus at zero strain loses its typical dependence on chirality. As soon as the tangent modulus is analyzed at non-zero tensile strain, this dependence is restored.

\subsection{Finite Elements}
\label{ssec_FEM} 
With the trained PICINN taking the role of a constitutive model, the finite element method can be readily applied. For this purpose, the FE code Abaqus was used, more specifically T3D2 truss/rod finite elements. Each truss finite element represents a single SWCNT and can be used to model nanotruss structures made of SWCNT. The PICINN then communicates with Abaqus via the UMAT user subroutine, which provides the stress and tangential modulus for the current strain and chiralities. With such an approach, the robust nonlinear capabilities of Abaqus can be readily utilized. As will be shown in the Results and Discussion section, these include in particular the handling of large displacements, which are rarely considered in analyzes of this type. Although the input data for the material nonlinearity is provided by the NN thru mentioned UMAT subroutine, all other aspects of it are included in Abaqus. In addition, Abaqus includes the modified Riks algorithm required for analyzes with bifurcations, i.e. stability problems. This issue is rarely covered in the literature, but is an important feature as it will be demonstrated in Sec.~\ref{ssec_Ex2}.

However, there is an additional issue that deserves further attention. In the geometrically nonlinear regime, the truss FE in Abaqus are treated as incompressible. This means that in each calculation step, the initial cross-sectional area of the rod is recalculated to ensure isochoric behavior. The dataset at hand, on the other side, involves the true strain and the true stress, which requires knowledge of the current SWCNT diameter. Therefore, as mentioned above, an additional PICINN was trained that outputs the current diameter for a given true strain and chirality. The architecture of this network is identical to that of the NN described above in terms of the number of layers and their interconnections, the number of neurons, the activation functions, etc. The only difference is that it is not trained on the derivatives like the previous one, but on its original form. This information was also passed to the UMAT subroutine in Abaqus and used to obtain the correct cross-sectional area. The implementation based on FE turned out to be robust and can reproduce MD simulations with excellent accuracy, thus drastically reducing computational costs without sacrificing MD quality. For more details, the reader is refereed to \cite{canadija2024computational}.

\subsection{Optimization Procedure}
\label{ssec_Opti} 
The goal of this research is to choose an efficient optimization method that could utilize the new nanotruss elements to design metamaterials with tailored properties or to design a specific nanotruss structure. As explained in the discussion above, such an optimization framework must take into account both the nonlinear behavior of the material and large displacements, i.e. geometric nonlinearities. This is a challenging task for any optimization algorithm. A convenient way to test the performance of the different optimization algorithms is through the application of the Indago package \cite{indago2024}. Optimization methods used in the present research are briefly described below. 

\textbf{Nelder-Mead (NM) method}

One of the most simple yet effective pattern search methods is the Nelder-Mead (NM) method \cite{nelder1965simplex}. Although primarily designed as a local search method, it is often competent even when solving multi-modal optimization problems. The technique utilizes an $n$-dimensional polytope with $n+1$ vertices that explores the search space through reflection, expansion, contraction and shrink operations performed on the polytope. We decided to use the NM variant with adaptive parameters \cite{gao2012implementing}, which is more persuasive than the initial variant, especially for problems of higher dimensions.

\textbf{Particle Swarm Optimization (PSO)}

PSO is one of the first attempts at utilizing swarm intelligence to solve optimization problems \cite{kennedy1995particle}. Since its initial publication, the PSO technique has undergone many improvements and extensions, but we have settled on a basic variant that still offers good performance based on the simple ideas of particle swarm motion. We can consider the PSO technique population-based, and the size of the swarm is the basic parameter of the method, along with inertial, cognitive, and social factors, it offers the possibility of a balance between local and global search.

\textbf{Artificial Bee Colony (ABC) optimization}

ABC is another heuristic optimization method inspired by the food-seeking behavior of honeybee swarms \cite{karaboga2009comparative}. The swarm motion employed in the ABC is even simpler than in PSO. However, there are three groups of bees (employed bees, onlookers and scouts) in the colony and the selection of moving bees is more complex and includes global, two types of local, and random mechanisms. By randomly generating scout bees, this algorithm ensures better global search and restrains premature convergence.

\textbf{Squirrel Search Algorithm (SSA)}

Similar to ABC, SSA is also inspired by foraging, specifically, it mimics the behavior of flying squirrels while seeking food \cite{jain2019novel}. Three distinct squirrel motions serve to explore the search space, motivated by different tree types from which squirrels start the flight. Additionally, the motion model also includes aerodynamic effects in gliding. The balance between local and global search is achieved by simulating seasonal changes, slowing down squirrel motion in winter, and eventually reinitializing their positions at the end of each winter season.

\textbf{Bat Algorithm (BA)}

Echolocation is a well-known phenomenon that allows bats to fly blindly in a given environment and seek prey. A very simplified version of this phenomenon is employed as a base idea to design the Bat Algorithm for the optimization \cite{yang2012bat}. Each bat in the swarm moves according to a local random walk by considering the varying frequency and loudness of pulses they emit. The rate of pulse emission increases asymptotically to the referent value while loudness increases towards zero as the optimization progresses. 

\section{Results and Discussion}
\label{sec_Example}
To illustrate the behavior of the proposed framework, a series of five carefully selected optimization design tasks is presented. The first example deals with a simple problem and analyzes the performance of the chosen optimization methods. The formulation of this problem is used as a starting point for other examples which is then adapted to the new example. The second example deals with the optimal design of an energy trapping structure that exploits two stable equilibrium configurations. The third example considers an auxetic nanotruss by resorting to multiobjective optimization. The fourth one presents optimization of a cantilever nanotruss, and the fifth one is about the metamaterial with the maximum compressive strength. The variety of examples in this section is intended to provide answers to the following questions: could CNT really be used as building blocks for metamaterials with exceptional properties, how to perform the optimal design of CNT-based metamaterials, and which optimization methods are best suited for this task.

\subsection{Weight Minimization of a Simple Nanotruss}
\label{ssec_Ex1}
\begin{figure}
	\centering
	\includegraphics[scale=0.4]{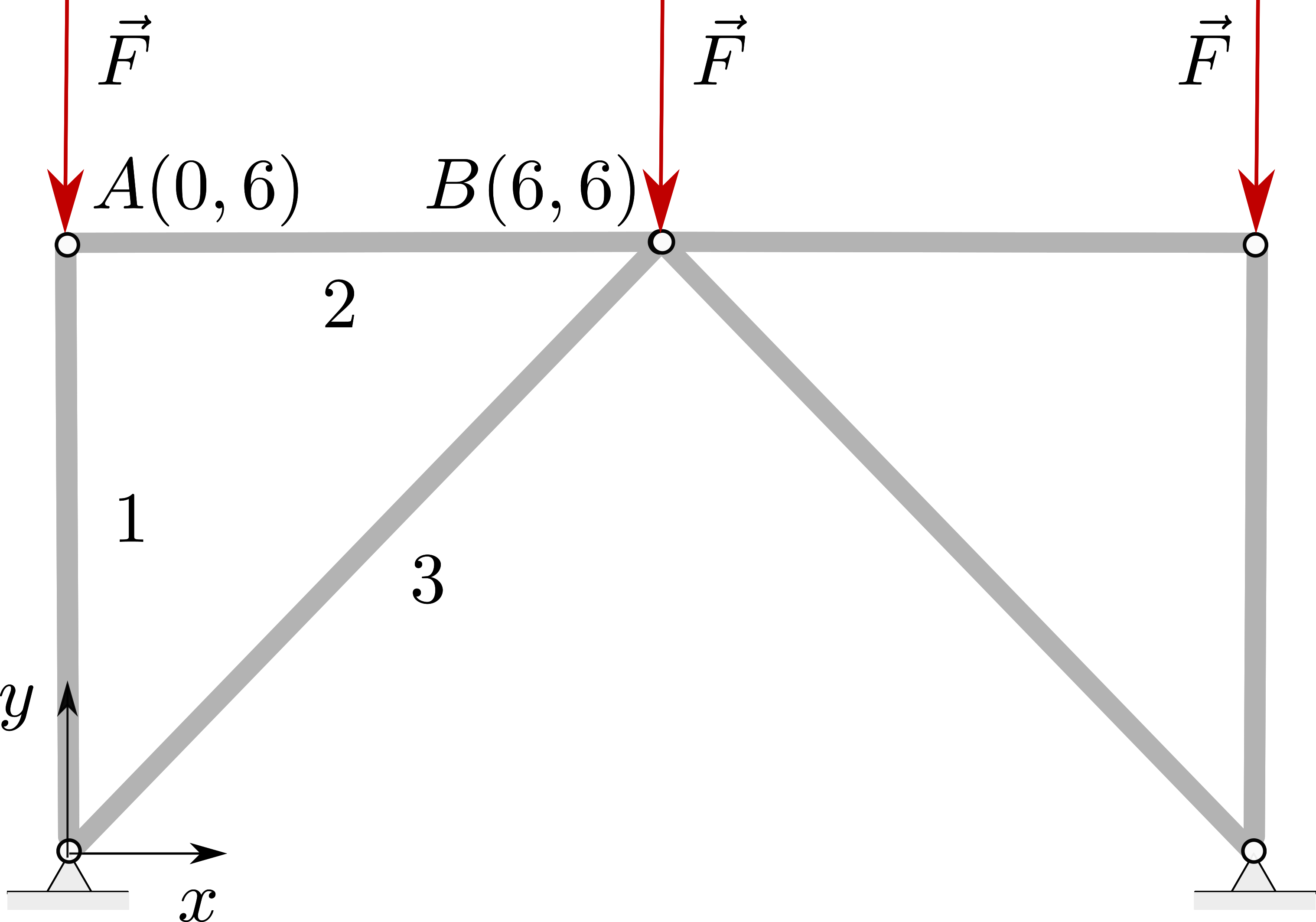}
	\caption{The initial configuration of a simple nanotruss.} 
	\label{fig_ex1_truss}
\end{figure}

The first example deals with the weight optimization of a simple nanotruss made of (10,0) SWCNTs, Fig.~\ref{fig_ex1_truss}. It is a very simple problem and this simplicity allows a straightforward interpretation of the results. This also enables a better insight into the behavior of optimization methods in CNT nanotrusses. To this end, the analysis is performed in detail in order to select the most suitable optimization method that could be used in other examples.

The initial diameter of the SWCNT rods as obtained from the uniaxial MD simulations is $D_0=0.783$ nm. The minimum (compressive) and maximum (tensile) stress at failure of the (10,0) CNT is $\sigma_\mathrm{c}=-69.68$ GPa and $\sigma_\mathrm{t}=84.81$ GPa respectively. The truss is supported and loaded as shown in Fig.~\ref{fig_ex1_truss}, where $F=50$ nN. Due to the symmetry of the problem, only half of the structure is discretized (rods 1, 2 and 3) and appropriate boundary conditions are applied. The overall dimensions of the symmetric model are $6\times6$ nm. The static analysis includes both material and geometrical nonlinearities due to the nonlinear stress-strain relationship and the large displacements. Moreover, SWCNT do not exhibit the same behavior in compression and tension, i.e. the stress-strain curves and consequently the tangent moduli $\mathrm{d} \sigma / \mathrm{d} \varepsilon =E_\mathrm{T}(\varepsilon)$ differ significantly in these regimes \cite{canadija2024computational}. The calculations were performed in Abaqus/Standard, using Newton method for handling/solving nonlinearity where the load is applied in 10 fixed step size increments.

The optimization problem can be summarized as follows:
\begin{equation}
	\label{eq_Ex1_opt_prob}
	\begin{array}{llc}
		\underset{\mathbf{x}}{\text{minimize}} & L(\mathbf{x})=L_1+L_2+L_3 & \\
		\text{subject to} &   \sigma_i(\mathbf{x})\le\sigma_\mathrm{t} & (\text{a}) \\
		&   \sigma_i(\mathbf{x})\ge\sigma_\mathrm{c} & (\text{b}) \\
		&   L_i(\mathbf{x}) \ge D_0 & (\text{c})\\
		&   L_i(\mathbf{x}) \le 12.5 D_0, \quad \text{if} \quad \sigma_i(\mathbf{x})\le 0 & (\text{d}) \\
		
	\end{array}	
\end{equation} 
where $\mathbf{x}=\lbrace x_A, y_A, y_B \rbrace $ are nodal coordinates and optimization variables, $L_i$ are the lengths of the rods and $\sigma_i$, $i=1,2,3$ are the true axial stresses in each rod element. The optimization problem is therefore to minimize the weight, which in this case is reduced to minimizing the total length $L$ of the rods. The total length of the rods in the initial configuration is 20.49 nm. Consequently, the task at hand requires changing the positions of the nodes $A(x_A,y_A)$, $B(6,y_B)$ during the optimization. The starting coordinates are $A(0,6)$ and $B(6,6)$. The abscissa $x_B$ of the point $B$ and the position of the supports are fixed and not subject to optimization. The nodal coordinates can take the values $x_A \in \left[ -15,15 \right]$, $y_A \in \left[ -9,21 \right]$ and $y_B \in \left[ -9,21 \right]$. In other words, the specified bounds for the optimization variables ensure that the nodal coordinates can change by $\pm 15$ relative to the initial positions $x_A, y_A, y_B$.

The above stated constraints can be explained in more detail. The limit values for the tensile stresses $\sigma_\mathrm{t}$ and the compressive stresses $\sigma_\mathrm{c}$ are used as stress constraints for each truss element ($\sigma_1$, $\sigma_2$, $\sigma_3$), (a, b). The constraint (c) enforces that the lengths $L_i$ of the unloaded truss elements must not be smaller than the initial diameter $D_0$. The last constraint ensures the validity of the shell-like buckling assumption at failure, which is observed in MD simulations. The length-to-diameter ratio of each member must be less than 12.5, i.e. $L_i\le9.78$ nm \cite{canadija2024computational}. The constraint (d) is active when the truss element is loaded in compression and is deactivated for members that are loaded in tension.

It is well known that the predictions of feed-forward neural networks, such as the one used here, can be of lower quality if the input is outside the range used in training. In this particular case, this happens when the strain becomes larger than strain at failure. Since the input variables for optimization can vary considerably, it is quite possible that the strain (and stress) in a rod is too large to obtain a reasonable prediction by NN. This will cause the FEM evaluation to fail and consequently the optimization process to fail. To solve the problem in this particular example, such evaluations are repeated, but now using the linearized behavior, i.e. the constant tangential stress-strain modulus is taken for $\varepsilon=0$. In this way, the evaluation will be successful, but the stresses will be overestimated and the constraints will be violated, as should be the case for such large strains. This ensures that the optimization process can continue even in such situations.

Five different optimization methods are tested, which are listed in Tab.~\ref{tab_Ex1_stat}. Majority of utilized optimization methods are stochastic and population based, and they start from randomly generated design vectors (for each individual of the population or swarm). On the other hand, the Nelder-Mead method is a local search that uses a simplex consisting of $n+1$ points (design vectors) in the $n$-dimensional search space. In order to ensure the robustness of the Nelder-Mead method, a simplex shape should not be degenerated \cite {lagarias1998convergence}. Hence, Indago initializes an orthogonal simplex from a single initial point by adding $n$ points displaced in each dimension by $\Delta x$. To compensate for the diversity of the population based initialization, a starting point for all methods is chosen as the best of 20 randomly generated designs. The analysis was performed until the maximum of \numprint{1000} evaluations or when 100 stalled evaluations are reached. The population/swarm size was 10 for ABC, BA, PSO and SSA. 

\begin{figure}
	\centering
	\includegraphics[scale=0.5]{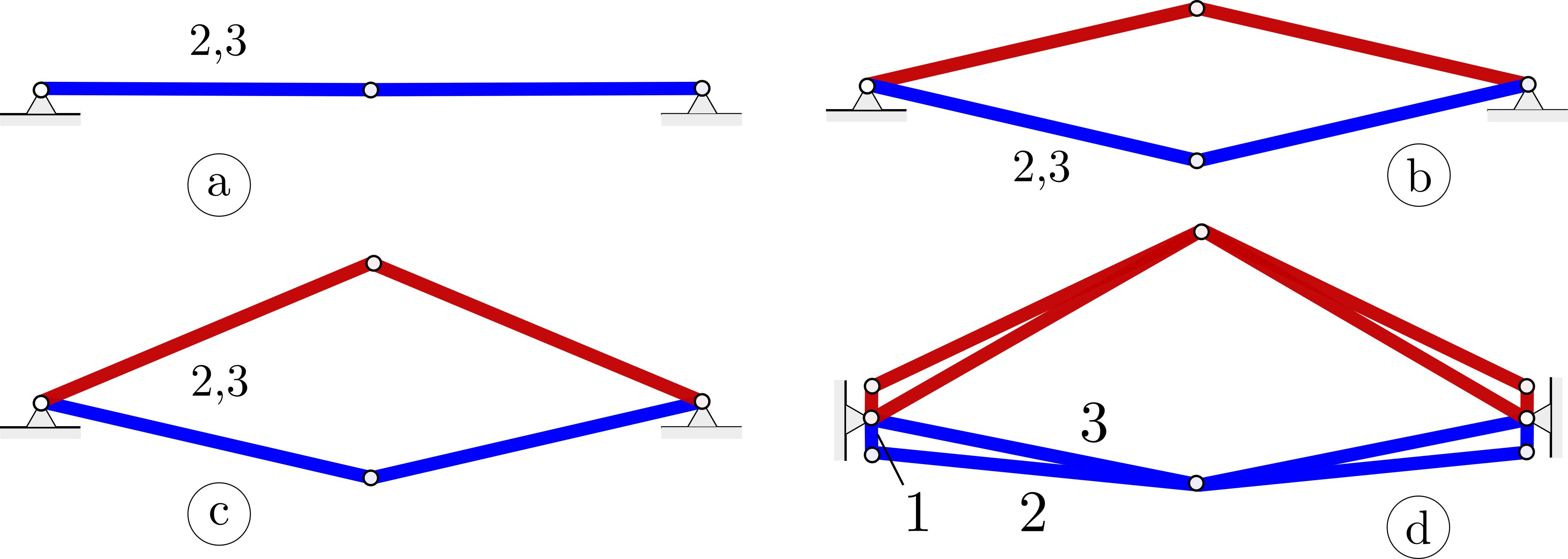}
	\caption{Schematic representations of local and global minimal configurations for various constraints: (a) global minimal configuration for unconstrained minimization; (b) two identical global minima for the constraint $\sigma_\mathrm{c} \le \sigma_i \le \sigma_t$, for $\left| \sigma_c\right| = \left| \sigma_\mathrm{t} \right|$; (c) a local (upper, red) and a global (lower, blue) minima for the constraint $\sigma_\mathrm{c} \le \sigma_i \le \sigma_\mathrm{t}$, for $\left| \sigma_\mathrm{c}\right| < \sigma_\mathrm{t}$; (d) a global (upper, red) and a local minimum (lower, blue) for the constraint used in (c) together with the additional constraint $L_i\ge D_0$. For the minimum it must be: $L_1=D_0$.} 
	\label{fig_ex1_ideal}
\end{figure}

To better interpret the results, first consider four simpler scenarios, Fig.~\ref{fig_ex1_ideal}. In order to keep the elaboration simple, all nonlinearities are neglected. In particular:
\begin{enumerate}[label=(\alph*)]
	\item In the unconstrained minimization, rod 1 will vanish, while rods 2 and 3 will be co-linear at the optimum. The forces in rods 2 and 3 will be infinite, Fig.~\ref{fig_ex1_ideal}a.
	\item If the stress constraint is introduced with the same absolute values used for the stress limits in tension and compression, there will be two configurations with the same value of the global minimum, Fig.~\ref{fig_ex1_ideal}b.
	\item If the absolute values of the allowable compressive stresses are lower than the allowable tensile stresses, then due to higher allowed values of the stresses, the configuration with tensile stresses involves smaller angles of the rods and consequently the shorter rods. The latter configuration then corresponds to the global minimum, Fig.~\ref{fig_ex1_ideal}c.
	\item The introduction of the constraint that the length of a rod cannot be smaller than the rod's initial diameter, Fig.~\ref{fig_ex1_ideal}d, means that rod 1 can no longer disappear and the lower configuration ($y_A < 0, y_B < 0$) shown is the global optimum with the tensile stresses in all rods. The upper configuration ($y_A> 0, y_B> 0$) is the local optimum with compressive stresses in all rods. Finally, since the last constraint $L_i \le 12.5 D_0$ serves as an upper bound for the lengths of the compressed rods, it reduces the feasible region of the domain but has no influence on the global optimum, so that the solution of the present problem should be similar to that in Fig.~\ref{fig_ex1_ideal}d.
\end{enumerate}

The statistical analysis of the various optimization and nanotruss model indicators is provided in Tab.~\ref{tab_Ex1_stat}. The lowest median of the objective function is obtained for the PSO, while the median for SSA is slightly lower. The ABC, BA and NM clearly show the lower performance. The standard deviation observed for PSO is at least a factor of two lower than for the BA, NM and SSA, indicating more consistent results. However, as mentioned above, multiple optima exist. It is interesting to note that only PSO always find a solution in the region of the search space that either includes the global or local optimum. Further, and very importantly, two almost identical lowest values of the objective function are provided by the NM, both for the global minimum (tensile stresses only and $y_A<0$,$y_B<0$) and for the local optimum (compressive stresses only and $y_A>0, y_B > 0$). The solution sgn $y_A\ne$ sgn $y_B$ is the case when the nodes $A,B$ are positioned on different sides of the horizontal line connecting the supports. Otherwise, NM is not as reliable as PSO as it manages to find only 5 global optima out of 10 runs. The worst optimization method for the problem at hand is BA, which manages to find only 1 global minimum in 10 runs, while the global minimum for PSO and SSA is found in 8 out of 10 runs. Looking at the number of evaluations until the optimization stops, ABC and BA also suffer from a stalled optimization process. This can also occur with other methods. We also present how many solutions out of 10 optimization runs are not feasible. While ABC and BA again perform the worst, NM, which provided the lowest global minimum, also fails to find feasible solutions in two runs, and SSA in one.  PSO always managed to find a feasible solution. Overall, 10 out of 50 solutions failed to satisfy all constraints. A selection of 12 solutions is documented in Tab.~\ref{tab_Ex1_optim}. To conclude, it can be said that the NM provides the lowest minima, even if the lower reliability requires more optimization runs. The PSO is the most reliable, but cannot outperform the optimum found by NM. 

\begin{table}
	\begin{center}
		{		\footnotesize
			\begin{tabular}{c|c|c|c|c|c}
				\hhline{|=|=|=|=|=|=|}
				Optimization method & ABC & BA & NM  & PSO & SSA  \\
				\hline
				Objective $L(x)$ & & & & &  \\
				\hline
				median			  & 17.622  & 19.044 & 17.721 & \underline{13.786}  & 14.045 \\
				average 	      & 19.114  & 23.502 & 17.874 & \underline{14.810}  & 16.438 \\
				st. dev.		  &  3.744  &  9.273 & 5.135  & \underline{2.606}	& 5.222  \\
				$L_\mathrm{min}$  & 15.403  & 13.324 & \underline{12.902} & 13.015 & 12.965  \\
				$L_\mathrm{max}$  & 26.570  & 40.028 & 28.759 & \underline{19.800} & 28.671  \\
				\hline
				Minimum type & & & & &  \\
				\hline					
				$y_A<0$,$y_B<0$ (global), 	  &  6 & 1 & 5 & 8 & 8 \\
				$y_A>0$,$y_B>0$ (local),	  &  2 & 2 & 4 & 2 & 1 \\
				sgn $y_A\ne$ sgn $y_B$ 		  &  2 & 7 & 1 & 0 & 1  \\
				\hline
				Not feasible 	& 4 & 3 & 2 & 0 & 1 \\	
				\hline
				Distance to optimum & & & & &  \\
				\hline
				median			  & 7.010 &  6.928 & 7.610 & \underline{3.462} & 3.692  \\
				average 	      & 7.148 &  8.591 & 6.516 & \underline{4.021} & 4.786  \\
				st. dev.		  & \underline{2.345} &  4.302 & 5.346 & 3.702 & 4.792  \\
				min  		      & 4.061 &  0.902 & \underline{0.000} & 0.399 & 0.174 \\
				max				  & 11.865 & 14.295 & 13.853 & \underline{10.235} & 13.805 \\			
				\hline						
				Evaluations & & & & & \\
				\hline				
				median  	    & 240  & 355 & 878  & 890  & 1010 \\
				average  	    & 260  & 390 & 808  & 755  & 860  \\
				min  			& 125  & 150 & 334  & 160  & 430  \\
				max  			& 455 & 670 & 1004 & 1010 & 1010\\					
				\hhline{|=|=|=|=|=|=|}
			\end{tabular}
		}	
		\caption{Statistical data on the performance of the optimization methods used - objective function, number of minimum types achieved, number of configurations achieved that were not feasible, distances to the global optimum and number of evaluations. Minimum values are underlined.}
		\label{tab_Ex1_stat}
	\end{center}
\end{table}

\begin{figure}
	\centering
	\includegraphics[scale=0.5]{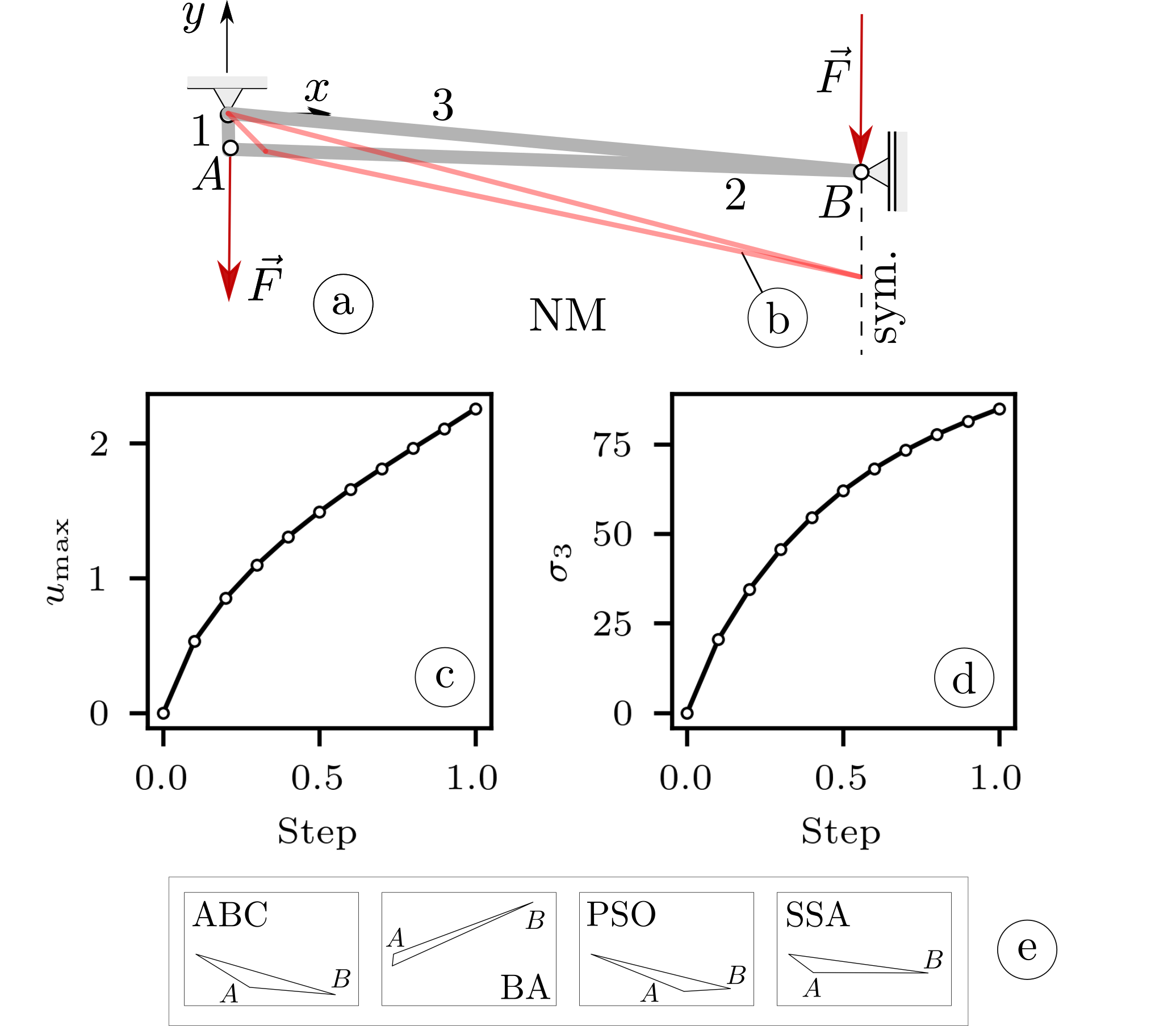}
	\caption{(a) Optimal configuration determined with the NM algorithm; (b) deformed configuration of (a); (c) evolution of the maximum displacement at point $B$; (d) evolution of the axial stress in rod 3; (e) sample configurations determined with other optimization methods.} 
	\label{fig_ex1_confs}
\end{figure}

Thus, it is clear that the NM algorithm, due to its local search abilities, provided the lowest value of the objective function. For this reason, the NM global minimum is used as a reference when discussing the mechanical performance. The solution shows the same behavior as in the simpler linear case initially discussed above - the length of rod 1 is reduced to the minimum allowable length, that is $L_1=D_0$. As already explained, due to the differences between the allowable tensile and compressive stresses, the minimum corresponding to the tensile stresses is also the global minimum. The maximum displacement was 2.254 nm, and the deformed configuration is shown in Fig.~\ref{fig_ex1_confs}a. The displacements are obviously large and their evolution is nonlinear, Fig.~\ref{fig_ex1_confs}c. The latter also applies to the stresses, Fig.~\ref{fig_ex1_confs}d. A small selection of optimal configurations for other optimization methods considered are shown in insets for comparison in Fig.~\ref{fig_ex1_confs}e.

\subsection{Design of a Energy Trapping Metamaterial}
\label{ssec_Ex2}
\begin{figure}
	\centering
	\includegraphics[scale=0.35]{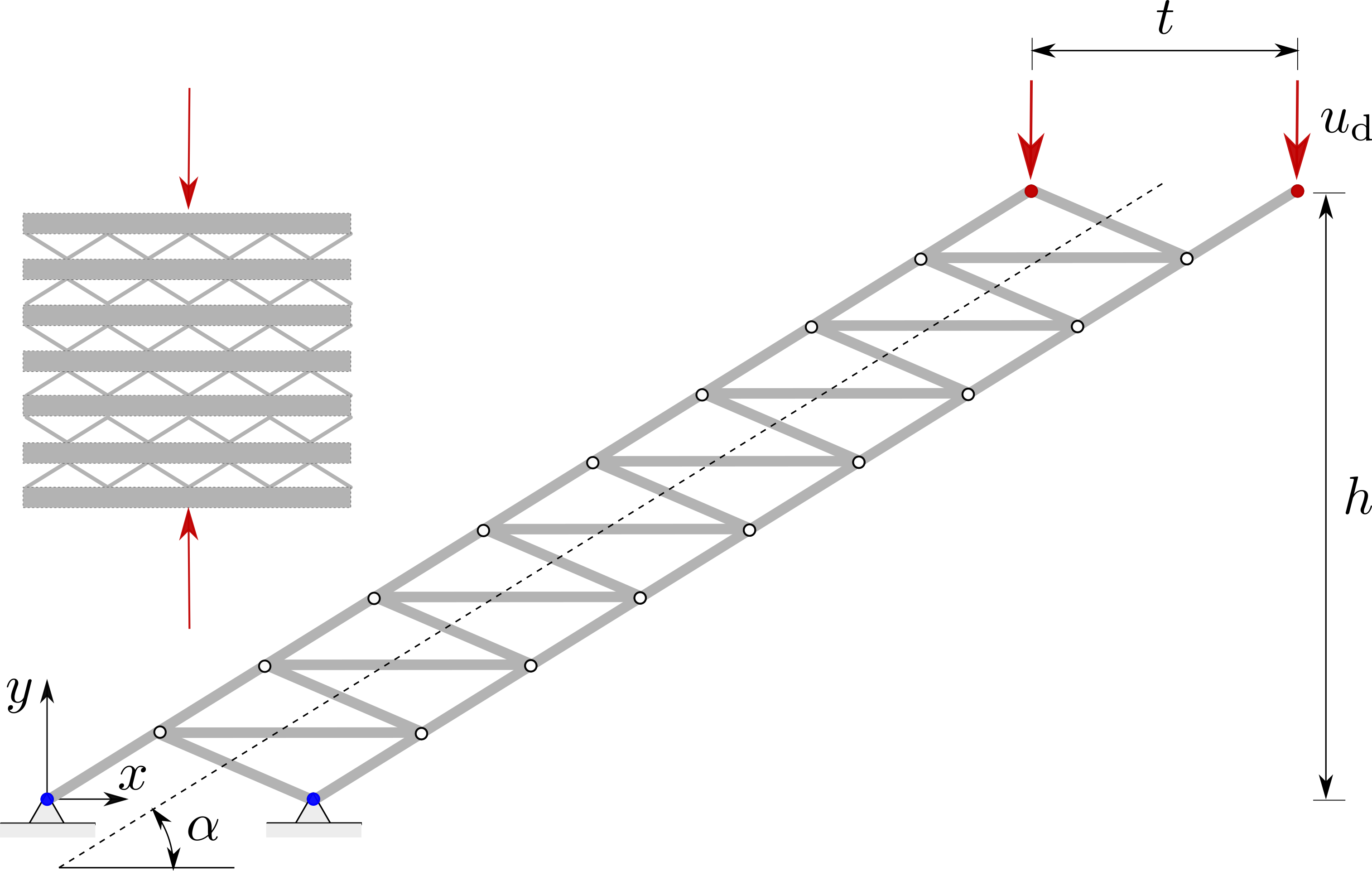}
	\caption{The nanotruss configuration used for energy trapping. The inset shows the repeating pattern used to construct a metamaterial; the straight rods schematically represent the nanotruss. Displacements $u_\mathrm{d}$ are prescribed in red colored nodes, supports in blue nodes. The ordinates of all other nodes represent the elements of the set $Y$.}
	\label{fig_ex2_nanotruss}
\end{figure}

The present example is motivated by the research presented in \cite{shan2015multistable}, in which an architected material was developed using tilted flexible beams fitted between two rigid layers. The polydimethylsiloxane beams were 1-6 mm long, symmetrically arranged, and repeated to form a new type of material, see the inset in Fig.~\ref{fig_ex2_nanotruss}. The beams were loaded in compression by prescribing displacements on one rigid layer while the other rigid layer was fixed. Since the loading direction did not coincide with the longitudinal axis of the beam, the beams were subjected to simultaneous bending and compression, resulting in a loss of stability at a critical point. While beams oriented perpendicular to the two rigid layers (i.e. vertically - in the loading direction) buckle elastically but cannot assume the second stable configuration and retain the initial configuration after the compressive load is removed, this is not the case for tilted beams. The latter may snap into a second stable configuration locking a part of the energy. They are referred to as bistable beams. When the load is reversed, the initial configuration can be fully restored. The first stable configuration is always the initial configuration, i.e. a straight, unloaded and tilted beam. In the second configuration, the beam shape resembles a slightly distorted and rotated letter "S", which happens after the bifurcation. In contrast to \cite{portela2021supersonic}, the deformation process is not associated with permanent damage to the material. Finally, it should be noted that in \cite{shan2015multistable} no optimization was performed, but the best result was selected from a series of calculations. 

The main property of such metamaterials is that they can absorb and trap the impact energy in a reversible manner as the energy of elastic deformation, thus offering a variety of practical applications. While in \cite{shan2015multistable} a beam is used to capture the energy, in the present application it is replaced by a planar nanotruss representing a basis of a metamaterial that exhibits similar behavior at the nanoscale.

The nanotruss was composed of (7, 0) SWCNT rods, with an initial diameter of $D_0=0.5564$ nm. The minimum (compressive) and maximum (tensile) stress at failure is $\sigma_\mathrm{c}=-107.97$ GPa and $\sigma_\mathrm{t}=81.07$ GPa. The nanotruss (Fig.~\ref{fig_ex2_nanotruss}) is simply supported at the bottom, while displacements $u_\mathrm{d}$ are prescribed at the other end.

Since there are two stable equilibrium configurations, the deformation process will not only involve large displacements, but also buckling at a certain point. To handle the bifurcation issue numerically, the modified Riks method was used, which is readily available in Abaqus. This is a significant complication in addition to the geometrically and materially nonlinear behavior of the truss. The inclusion of this example in the present research thus demonstrates the framework's ability to cope with numerically demanding problems such as buckling. 

In the case of a failed FE analysis there is no solution for the given load, so the absolute value of the stresses at failure multiplied by 10 is set for all stress constraints. The same procedure is used in the remaining examples. 

Three variables $\mathbf{x}=\lbrace\alpha,h,t\rbrace$ are taken into account in the optimization, i.e. the angle $\alpha \in \left[ 10\circ,90\circ \right]$ that the nanotruss makes to the horizontal line, the height of the truss $h \in \left[ 3, 60 \right]$ nm and the width at the horizontal cross-section $t\in \left[ D_0, 2/3L_\mathrm{max} \right]=\left[ 0.5564, 4.6362 \right]$ nm, Fig.~\ref{fig_ex2_nanotruss}. The maximum length of the individual rods is defined as $L_\mathrm{max}=12.5D_0=6.9544$ nm to exclude a beam-like buckling, as explained in Ex.~\ref{ssec_Ex1}.

\begin{figure}
	\centering
	\includegraphics[scale=0.4]{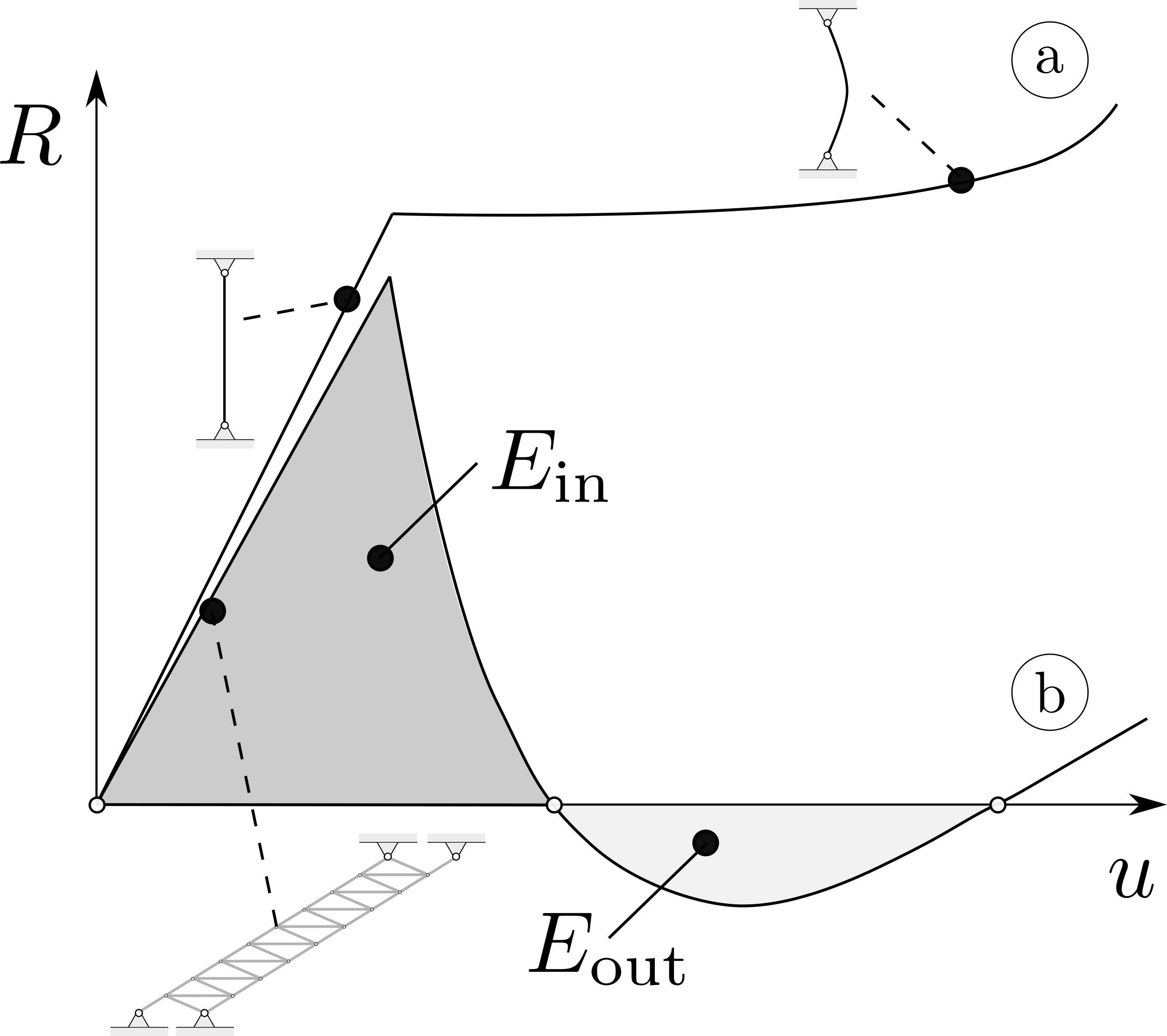}
	\caption{$R-u$ curves in the case of stability loss: (a) an energy non-trapping simply supported beam, (b) an energy trapping nanotruss. The rightmost zero point corresponds to the second stable configuration.} 
	\label{fig_ex2_energy}
\end{figure}

Since the purpose of the present metamaterial is to store the deformation energy, the energy aspects of the problem should be described in more detail. A schematic representation of the energy parts involved in the deformation process is given in Fig.~\ref{fig_ex2_energy}. The second stable equilibrium point corresponds to the rightmost zero point in the curve (b). The total energy at this point is then given as
\begin{equation}
	\label{eq_Ex2_energy}
	E_\mathrm{tot}=	E_\mathrm{in}-E_\mathrm{out}= \int R(u) \mathrm{d} u
\end{equation} 
where the notation is explained in Fig.~\ref{fig_ex2_energy}. $R$ is the reaction force in the supports, and $u$ is the prescribed displacement at the current time increment where $0 \le u \le u_\mathrm{d}$. Therefore, in the structure that has assumed the second stable configuration, the first configuration can be recovered by introducing energy $E_\mathrm{out}$ into the system.

The main idea in this example is to develop a metamaterial that can store as much energy $E_\mathrm{tot}$ as possible. The optimization problem can now be defined as minimization:
\begin{equation}
	\label{eq_Ex2_opt_prob}
	\begin{array}{llc}
		\underset{\mathbf{x}}{\text{minimize}} & -E_\mathrm{tot}(\mathbf{x}) & \\
		\text{subject to} &   \max S \le\sigma_\mathrm{t} & (\text{a}) \\
		&    \min S \ge \sigma_\mathrm{c} & (\text{b}) \\
		&   L_{n_\mathrm{e}} \ge D_0 & (\text{c})\\
		&   L_{n_\mathrm{e}} \le 12.5 D_0, \quad \text{if} \quad \sigma_{n_\mathrm{e}}(t_i)\le 0 & (\text{d}) \\
		&   E_\mathrm{out} \ge 10^{-6} & (\text{e}) \\
		&   Y \ge 0, & (\text{f}) \\
		&   Y \le y_\mathrm{d}, & (\text{g}) \\
	\end{array}	
\end{equation} 
where $S$ is a set containing the stresses $\sigma(n_\mathrm{e},t_i)$ in all rods $n_\mathrm{e}=1,..N_\mathrm{e}$ calculated for all load increments $t_i,i=1,..,N$. Above, $N_\mathrm{e}=39$ and $N$ represent the number of rods in the nanotruss and the number of increments used in the calculation, respectively. The set $Y$ contains the vertical coordinates $y_{n_\mathrm{n}}(t_i)$, $n_\mathrm{n}=1,..N_\mathrm{n}$ of all nodes in the deformed structure with the exception of the nodes at which the displacements are prescribed ($u_\mathrm{d}$ or supports), and $N_\mathrm{n}=22$ is the number of nodes in the nanotruss. The vertical coordinate of the nodes in the deformed structure where displacements are prescribed is $y_\mathrm{d}(t_i)$. 

The objective function in Eq.(\ref{eq_Ex2_opt_prob}) thus refers to areas under the $R-u$ curve, which are evaluated from a series of discrete points at each increment. The area is obtained by applying the composite trapezoidal rule. Note that the last point used in the calculation is the second stable configuration point in Fig.~\ref{fig_ex2_energy}. However, the modified Riks procedure does not use exactly this point, so the values for $R, u$ are linearly interpolated from two nearest increments. Thus, the optimization problem is to maximize the stored energy ($E_\mathrm{tot}$), while the constraint Eq.(\ref{eq_Ex2_opt_prob})$_{e}$ requires that the energy required for the nanotruss to snap back to the undeformed configuration ($E_\mathrm{out}$) exists, which is conveniently enforced by using a small number ($10^{-6}$). Of course, some more stable systems can be obtained by requiring larger $E_\mathrm{out}$.

The constraints regarding the stresses in Eq.(\ref{eq_Ex2_opt_prob})$_{a,b}$ are much more difficult to handle than those used in Ex.~\ref{ssec_Ex1}. In particular, the stress evolution in the former example is a monotonically increasing function, so it is sufficient to monitor the last calculated value to ensure that the stress constraints are fulfilled. In the present example, this is not the case as the stresses change abruptly when the bifurcation occurs. For this reason, a set of stresses determined for all time increments and for all rods is obtained, and the maximum and minimum stresses from this set are used in Eq.(\ref{eq_Ex2_opt_prob})$_{a,b}$. Only these two values are constrained to be within the feasible domain.

The constraints for the length of the rods Eq.(\ref{eq_Ex2_opt_prob})$_{c,d}$ are the same as in Ex.~\ref{ssec_Ex1}. The remaining constraints Eq.(\ref{eq_Ex2_opt_prob})$_{f,g}$ ensure that the ordinate of any node does not go below the ordinates in which the supports are placed ($y=0$) and do not go above the ordinates $y_\mathrm{d}$ of nodes in which the displacements are prescribed. This means that the nodes between two ends must always remain between the two ends during deformation to avoid contact between the nanotruss and the rigid layers.

The NM, SSA and PSO optimization methods were selected with the same parameters as in Ex.~\ref{ssec_Ex1}. The performance indicators are documented in Tab.~\ref{tab_Ex2_stat}. All optimization methods were run 10 times. This should be sufficient to demonstrate the main ideas of the present research, although a more rigorous optimization might require a larger number of runs. Comparing the statistical indicators with those in Tab.~\ref{tab_Ex1_stat} provides conclusions that these are not so different from those in Ex.~\ref{ssec_Ex1}. This discussion can be found at the end of the example.

\begin{figure}
	\centering
	\includegraphics[scale=0.25]{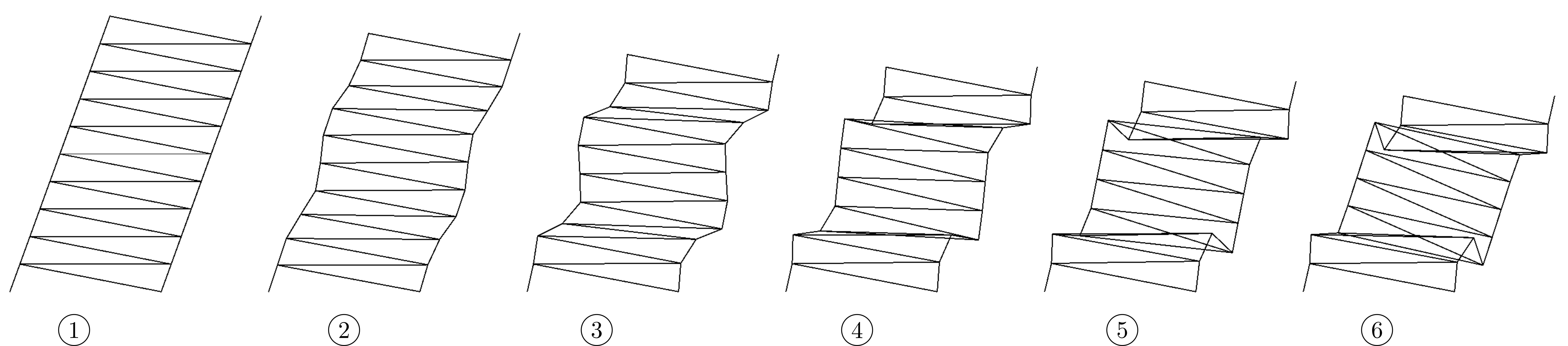}
	\caption{Deformed shapes of an energy trapping nanotruss for a global optimal configuration; the numbers indicate the corresponding points in Fig.~\ref{fig_ex3_RU}. Configuration (6) is the second stable configuration, while the unstable equilibrium is shown in (4).} 
	\label{fig_ex4_global}
\end{figure}

\begin{figure}
	\centering
	\includegraphics[scale=0.25]{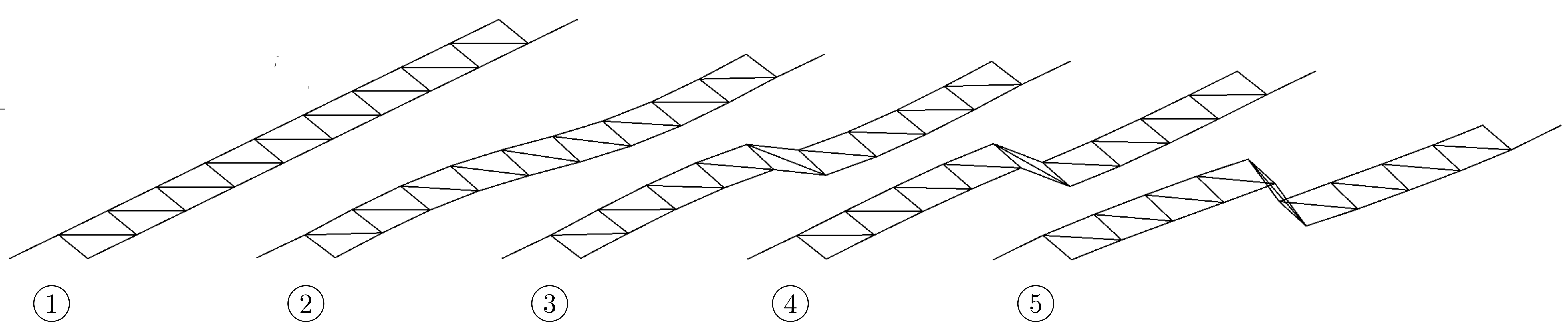}
	\caption{Deformed shapes of an energy trapping nanotruss for a local optimal configuration; the numbers indicate the corresponding points in Fig.~\ref{fig_ex3_RU}, cf. the results in \cite{shan2015multistable}. Configuration (5) is the second stable configuration, while the unstable equilibrium is not shown due to its proximity to configuration (5). } 
	\label{fig_ex5_local}
\end{figure}

\begin{figure}
	\centering
	\includegraphics[scale=0.4]{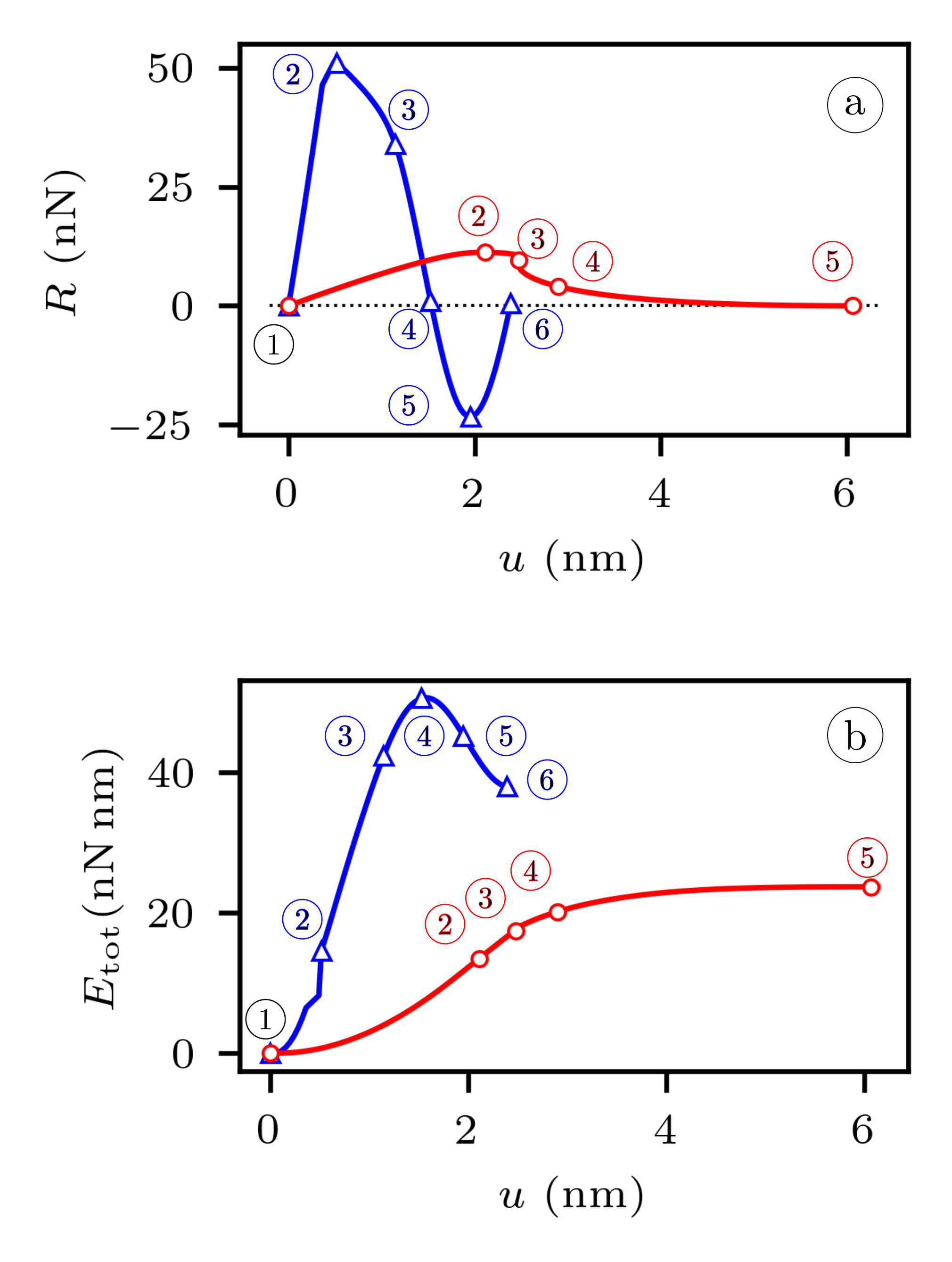}
	\caption{(a) $R-u$ curves for the global (blue, circles) and local (red, triangles) optimal configuration, (b) $E_\mathrm{tot}-u$ curves for the global and local configuration; the numbers indicate the configurations shown in Figs.~\ref{fig_ex4_global}\&\ref{fig_ex5_local} } 
	\label{fig_ex3_RU}
\end{figure}

It turned out that the optimizations converge to two minima (Tab.~\ref{tab_Ex2_optim}), one global corresponding to a more complicated deformation shape (Fig.~\ref{fig_ex4_global}) with a trapped energy of 37.980 nN$\,$nm and the second, lower one (herein will be referred to as the local minimum) with a simpler deformation shape (Fig.~\ref{fig_ex5_local}) and a trapped energy of 23.709 nN$\,$nm. The two configurations differ significantly with respect to the angle $\alpha$: the global optimum corresponds to a more tilted nanotruss at $\alpha{\sim}70\circ$, and with a thickness-to-height ratio of 0.549. The evolution of the total reaction force and total energy for both cases is shown in Fig.~\ref{fig_ex3_RU}. For the local optimum, $E_\mathrm{out}$ is close to zero, so it is difficult to notice the part of the curve below the axis $R=0$, Fig.~\ref{fig_ex3_RU}a. In the case of the global optimum, however, this part of the curve is clearly visible. The situation is similar with Fig.~\ref{fig_ex3_RU}b, where the drop in the local optimum curve cannot be visually noticed.

The local minimum configuration corresponds to $\alpha{\sim}26\circ$, $t/h=0.327$. Remarkably, \cite{shan2015multistable} reports the absence of the second stable configuration for beams having almost the same value, $\alpha{\sim}25\circ$ and below, i.e. the beams snap back to the initial configuration when unloaded. Clearly, this is characterized by the absence of $E_\mathrm{out}$. For the present local minimum nanotruss configuration at $\alpha{\sim}26\circ$, $E_\mathrm{out}=1\cdot10^{-6}$, i.e. it is almost zero, which is to be expected at the point where $E_\mathrm{tot}$ reaches the maximum. 

The overall dimensions of the nanotruss, excluding the rigid horizontal parts used to introduce the load, are 7.531$\times$8.257 nm for the global configuration and 33.593$\times$14.166 nm for the local configuration. For the best global and local configuration in Tab.~\ref{tab_Ex2_optim}, this gives a specific absorbed energy, expressed over the total area occupied by the truss as 0.696 J/m$^2$ for the global and 0.0498 J/m$^2$ for the local configuration, for in-plane loading of a single layer of repetitive nanotrusses. 

The deformation shape corresponding to the global minimum is manifested by the rotation of the central part of the truss which rotates between two truss sections. In contrast to the global minimum, configurations near the local minimum resulted in the simple shape with a kink that forms approximately in the midspan of the nanotruss. The kink allows a redistribution of the normal stresses in the rods, and due to the balance of the moment of momentum, the stresses can now self-equilibrate without the need for external loading. This is the second stable configuration. A similar type of deformation also occurs in the case of a monolithic beam, \cite{shan2015multistable} as described earlier. Lastly, it is emphasized that the present case considers a planar structure, so that the neighboring parallel layers of the repeating pattern (inset in Fig.~\ref{fig_ex2_nanotruss}) should be connected to each other in a suitable way to avoid possible out-of-plane buckling.

The global and local minima are used as a reference and the distance to these as a measure of how close the other results are. Due to the different nature of the variables, i.e. two linear dimensions and one angular, the variables were normalized before the distance was calculated. The min-max feature scaling $x_\mathrm{norm}=(x-x_\mathrm{l})/(x_\mathrm{u}-x_\mathrm{l})$ is used for each variable, where the upper $x_\mathrm{u}$ and lower $x_\mathrm{l}$ bounds on variables are used. The results are considered close to one of the two optima if the normalized distance is less than 0.10. It is immediately apparent that although 30 optimizations were performed, only 2 global optima were detected, Tab.~\ref{tab_Ex2_optim}. Compared to Ex.~\ref{ssec_Ex1}, the detection of local optima is also significantly worse - only 10 were found. This is a strong indication that this is a much more challenging problem. 

As far as optimization methods are concerned, SSA was able to find 5 optima and was also the only method that found the global minimum. In contrast, Nelder-Mead was the only optimization method that managed to find the two best local minima, although it did not find a global minimum in 10 runs. PSO found slightly more optima, but also failed to find a global minimum. The latter two optimization methods are also attractive in terms of computational costs, as they require a median of about 500 evaluations, while NM never reaches the maximum number of evaluations. An SSA and an NM solution were not feasible.

\begin{table}
	\begin{center}
		{		\footnotesize
			\begin{tabular}{c|c|c|c}
				\hhline{|=|=|=|=|}
				Optimizer: &  NM  & PSO & SSA   \\
				&     &     &      \\
				\hline
				Objective $-E_\mathrm{tot}(x)$ & & &\\
				\hline
				median			      &  3.808 & 16.309 & 20.593  \\
				average 	          &  8.273 & 14.269 & 17.285  \\
				st. dev.	    	  & 10.848 &  8.422 & 13.866  \\
				$-E_\mathrm{tot,min}$  & 0.266 &  1.697 &  0.005  \\
				$-E_\mathrm{tot,max}$ & 23.709 & 23.707 & 37.980  \\				
				\hline
				Number of minima found  & 3 & 4 & 5  \\
				\hline			
				Evaluations & & &  \\
				\hline				
				runs	  	    & 10  & 10  & 10\\
				median  	    & 516  & 540  & 735 \\
				average  	    & 520  & 587  & 699  \\
				min  			& 424  & 130  & 150 \\
				max  			& 703 & 1010 & 1010 \\					
				\hhline{|=|=|=|=|}
			\end{tabular}
		}	
		\caption{Statistical data on the performance of the optimization method used - objective function, number of particular minimum types found and number of evaluations. One SSA and one NM solution were not feasible. Energy in nN$\,$nm. }
		\label{tab_Ex2_stat}
	\end{center}
\end{table}

\begin{table}
	\begin{center}
		\addtolength{\leftskip} {-2cm}
		\addtolength{\rightskip}{-2cm}
		{		\footnotesize
			\begin{tabular}{c|r|r|r|c|r|r|r|r|c}
				\hhline{=|=|=|=|=|=|=|=|=|=}
				opt. & $h$(nm) & $t$(nm) & $\alpha$($\circ$) & $E_\mathrm{tot}$ & $E_\mathrm{out}$ & $\sigma_\mathrm{min}$ & $\sigma_\mathrm{max}$&  eval. & End \\
				&  &   &   & (nN$\,$nm) 			    & (nN$\,$nm)			   &  (GPa) &  (GPa) &   &  \\
				\hline						
				SSA$^1$ & 8.257	&	4.534 &	70.048 &	37.980 &	1.27E+01 & -104.56 &	75.34 &	1010 &	M \\
				SSA		& 8.351 &	4.636 &	66.498 &	34.617 &	1.07E+01 &	-97.99 &	72.38 &	1010 &	M \\
				\hline
				NM$^2$	&14.166 &	4.636 &	26.068 &	23.709 &	1.00E-06 &	-45.56 &	81.07 &	526 &	S \\
				NM		&14.166 &	4.636 &	26.068 &	23.708 &	1.00E-06 &	-45.56 &	81.07 &	509 &	S \\
				PSO		&14.166 &	4.636 &	26.069 &	23.707 &	8.49E-06 &	-45.56 &	81.07 &	1010 &	M \\
				SSA		&14.167 &	4.636 &	26.109 &	23.659 &	5.72E-03 &	-45.58 &	81.05 &	1010 &	M \\
				PSO		&14.153 &	4.636 &	26.090 &	23.583 &	6.02E-03 &	-45.51 &	80.97 &	500 &	S \\			
				PSO		&14.105 &	4.619 &	26.092 &	23.533 &	4.69E-03 &	-45.54 &	81.01 &	720 &	S \\
				NM		&13.803	&   4.517 &	26.069 &	23.101 &	5.39E-06 &	-45.56 &	81.07 &	561 &	S \\
				SSA		&14.338 &	4.636 &	27.795 &	22.119 &	1.23E+00 &	-46.84 &	81.07 &	920 &	S \\
				SSA		&13.303 &	4.342 &	26.410 &	21.989 &	1.05E-01 &	-45.75 &	81.07 &	540 &	S \\
				PSO		&14.486 &	4.636 &	32.395 &	15.911 &	7.20E+00 &	-50.53 &	81.02 &	1010 &	M \\
				\hhline{=|=|=|=|=|=|=|=|=|=}
			\end{tabular}
		}	
		\caption{Selection of results. The first two rows correspond to the global minimum, the rest to the local minimum. Optimizers used (opt.), variables, energies, minimum and maximum stress in the rods, number of evaluations (eval.), type of optimization end: S - stalled evaluations, M - maximum number of evaluations reached; $^1$overall minimum objective (global), $^2$second minimum objective (local).}
		\label{tab_Ex2_optim}
	\end{center}
\end{table}

\subsection{Truss-Based Auxetic Material}
\label{ssec_Ex3}
This example considers the development of a material based on a two-dimensional auxetic nanotruss structure. The present design is a common choice in the development of auxetic materials, and its popularity is also the main reason for its selection for this study. However, for the first time, the structure of the material is assembled of parallel layers of nanotrusses, and each layer expands laterally when stretched along the loading axis. As a result, the Poisson's ratio $\nu$ will be negative with respect to these two axes. A single elementary cell of a layer will be analyzed in the present example.

Recall that the behavior of isotropic and anisotropic materials is fundamentally different in this respect, since the value of Poisson's ratio is restricted for isotropic solids. In the three-dimensional case, for example, the ratio can assume values of $-1\le\nu\le0.5$, and in the two-dimensional case it is $-1\le\nu\le 1$ \cite{landau1975,kw2003remarks}. On the other hand, there are no restrictions on $\nu$ for anisotropic materials. Auxetic materials have a negative Poisson's ratio and are rare in nature. Examples are pyrite with a Poisson's ratio of -1/7 \cite{love2013treatise}, CsH$_2$PO$_4$ $\nu_\mathrm{min}=-1.93$ and $\nu_\mathrm{max}=2.70$, and LaNbO$_4$ $\nu_\mathrm{min}=-3.01$ and $\nu_\mathrm{max}=3.95$ \cite{norris2006extreme}. For further examples see \cite{carneiro2013auxetic,baughman1998negative,li2023auxetic} and for carbon-based metamaterials \cite{cai2021hierarchical,wu2013fracture,pedrielli2017designing,cai2023tunable}.

\begin{figure}
	\centering
	\includegraphics[scale=0.30]{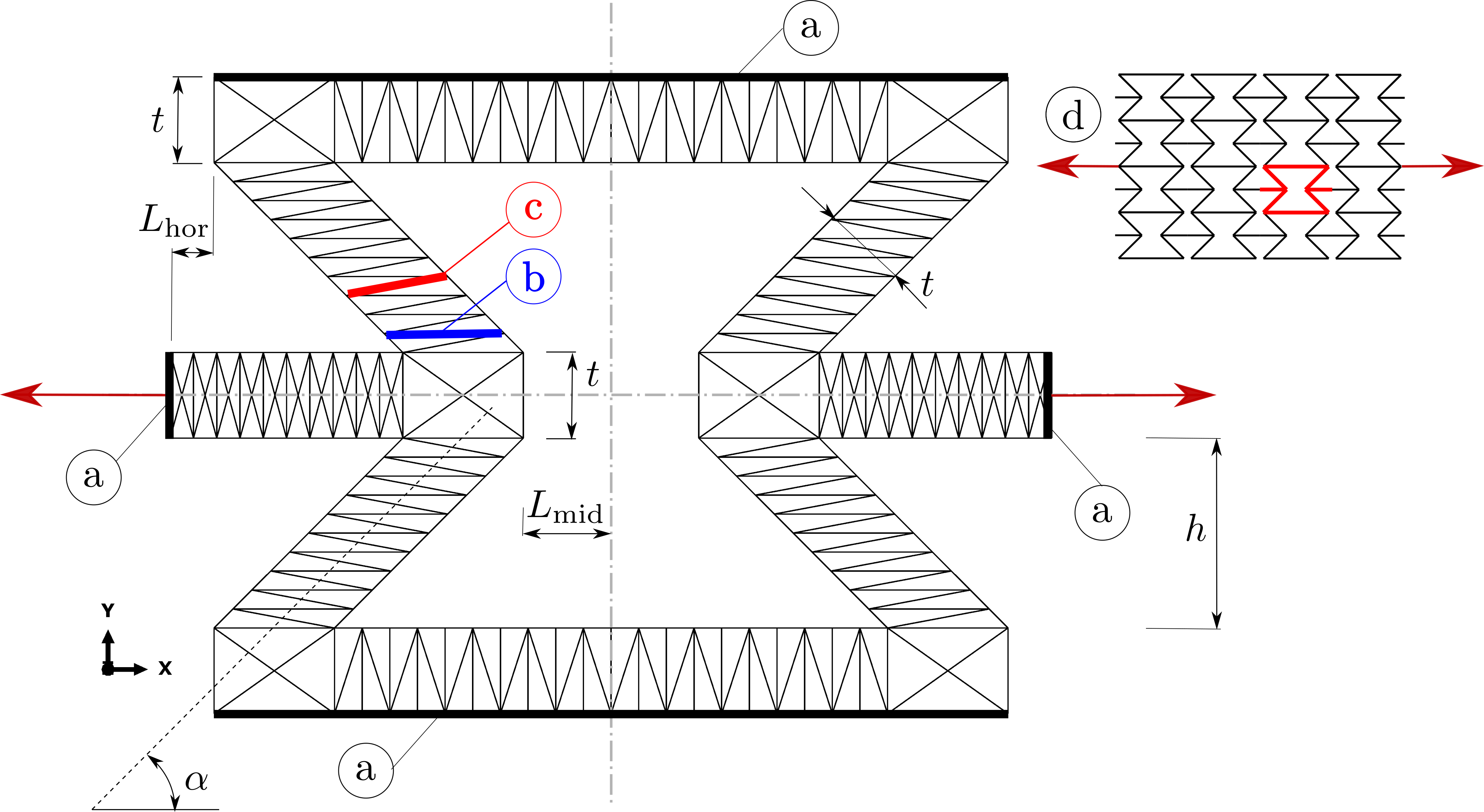}
	\caption{Elementary cell of the auxetic nanostructure with optimization variables. (a) edges with applied periodic boundary conditions (black); (b) rods loaded in compression (blue); (c) rods loaded in tension (red); (d) repeating pattern.} 
	\label{fig_ex3_aux_cell}
\end{figure}

When it comes to materials with engineered structures, anisotropy opens up further possibilities. From several possible choices \cite{fozdar2011three} a popular reentrant honeycomb configuration is chosen in this example, Fig.~\ref{fig_ex3_aux_cell}. This type of structure belongs to the class of stiff auxetics. One of the first attempts to analyze such a configuration, albeit in a three-dimensional context, was in the development of polymer foam, \cite{lakes1987foam}, see also \cite{ashby2006properties} for a general context of the mechanical behavior of foams. The closest approach to the present model of cell modeling is a general concept \cite{rayneau2018stiff}, in which it was presented without specialization to a particular material. In contrast to the rod/truss elements used here, the model contained therein was built from beams. Likewise, no attempt was made to optimize the design and the constant rib angle of 45$\circ$ was used. The model also considers the classical Euler beam-like buckling, which is not often encountered in other works. In the end, auxetic trusses sometimes use the node or beam bending stiffness \cite{desmoulins2016auxeticity,chen2021bistable}. In the present approach, the role of bending stiffness is taken not by individual beam elements, but by several nanotruss members that combined can take the bending load.

The initial nanotruss design is shown in Fig.~\ref{fig_ex3_aux_cell}. The structure is composed of (20, 20) SWCNTs. Based on \cite{canadija2021deep,canadija2024computational}, such nanotubes buckle at $\epsilon_\mathrm{c}=-0.0189$, $\sigma_\mathrm{c}=-19.533$ GPa, while the fracture in tension occurs at $\epsilon_\mathrm{t}=0.2267$, $\sigma_\mathrm{t}=132.061$ GPa. The initial diameter at 300 K is $D_0=2.660$ nm, so that the corresponding initial cross-sectional area is $A=2.839$ nm$^2$. Compared to the previous example, this nanotube configuration has a larger diameter and consequently fails by shell buckling at lower stress and strain.

The Poisson's ratio $\nu$ is calculated as usual:
\begin{equation}
	\label{eq_Ex3_Poisson}
	\nu= - \frac{\Delta L_y L_x }{ L_y \Delta L_x}.
\end{equation}
Here $\Delta L_x, \Delta L_y$ are the differences of the lengths in the horizontal ($x$, loading direction, Fig.~\ref{fig_ex3_aux_cell}) and vertical ($y$, perpendicular to the loading direction). The initial lengths in these directions are $L_x, L_y$. Note that due to the auxetic character of the structure, $\Delta L_y$ is positive, as are other terms, resulting in a negative Poisson's ratio. It is also clear that the Poisson's ratio will not be constant during deformation. As the rib rotates, the largest incremental perpendicular displacements take place at the beginning and decrease in a nonlinear manner with the deformation. The Poisson's ratio at the end of the deformation is used as the optimization goal as defined below. Note that once the rib has assumed the vertical position, lateral expansion ceases. Further longitudinal deformation would lead to a classic lateral contraction.

Now, due to the different orders of magnitude in $\nu$ and $L_\mathrm{tot}$, these terms must be rescaled for the application in the fitness function:
\begin{equation}
	\label{eq_Ex3_rescaled}	
	\overline{\nu}=1-\frac{\nu}{\nu_\mathrm{min}}, \quad 
	\overline{L}_\mathrm{tot}=\frac{L_\mathrm{tot}-L_\mathrm{min}}{L_\mathrm{max}-L_\mathrm{min}}.
\end{equation}
The indices max and min denote the maximum and minimum possible value of the Poisson's ratio and the total length. 

The structure is parameterized with the variables $\mathbf{x}=\lbrace h, t, \alpha, L_\mathrm{hor}, L_\mathrm{mid} \rbrace$, Fig.~\ref{fig_ex3_aux_cell}. The lower and upper bounds on variables are given in Tab.~\ref{tab_Ex3_optim}. The optimization problem is now defined as a multiobjective problem:
\begin{equation}
	\label{eq_Ex3_opt_prob}
	\begin{array}{llc}
		\underset{\mathbf{x}}{\text{minimize}} & \omega_1 \overline{\nu} (\mathbf{x}) + \omega_2 \overline{L}_\mathrm{tot} & \\
		\text{subject to} &   \max \sigma_{n_\mathrm{e}} \le\sigma_\mathrm{t} & (\text{a}) \\
		&    \min \sigma_{n_\mathrm{e}} \ge \sigma_\mathrm{c} & (\text{b}) \\
		&   L_{n_\mathrm{e}} \ge D_0 & (\text{c})\\
		&   L_{n_\mathrm{e}} \le 12.5 D_0, \quad \text{if} \quad \sigma_{n_\mathrm{e}}(t_i)\le 0 & (\text{d}). \\
	\end{array}	
\end{equation} 
Above, $\sigma_{n_\mathrm{e}}$ are the stresses in rods at increments $t_i$, $n_\mathrm{e}=1,..N_\mathrm{e}$, $N_\mathrm{e}$ is the number of rods, $\nu (\mathbf{x})$ is the Poisson's ratio and $L_\mathrm{tot}(\mathbf{x})$ is the total length of all rods in the nanostructure, and $\omega_1=0.30$ and $\omega_2=0.70$ are weight factors. The constraints remain in the line with previous examples, i.e. constraining stresses and lengths within certain limits. The fitness function requires additional elaboration. Obviously, the goal is to obtain a nanostructure that favors larger negative Poisson ratios while simultaneously keeping the length (or mass) as small as possible. Note that other objectives could be added to the current fitness function. For example, the maximum stiffness in the loading direction and in the lateral direction could be part of the formulation.

To completely define the above optimization problem, it remains to describe how to obtain maxima and minima in Eq.(\ref{eq_Ex3_rescaled}). Since the central idea of the above rescaling is to bring both terms to the same order of magnitude, it is not necessary to obtain exact maximum and minimum values, but only a rough approximations. In particular, $\nu_\mathrm{min}=-17.825$ was obtained as a minimum of the similar optimization problem to Eq.(\ref{eq_Ex3_opt_prob}), but with the fitness function replaced by $\underset{\mathbf{x}}{\text{minimize}} \thickspace \nu (\mathbf{x})$, i.e. in the single objective optimization. Five runs were performed and the minimum Poisson's ratio was selected. The total lengths $L_\mathrm{max}=24181.0$ nm and $L_\mathrm{min}=762.1$ nm were determined for structures defined with minimum and maximum bounding values of the design vector. 

Periodic boundary conditions were defined at the boundary of the nanostructure. The movement of the nodes on the left side has been restricted, while the nodes on the right side have been moved horizontally by $\Delta L_x= 1.1 h/\tan(\alpha)$, which means that the structure will be extended slightly more than the horizontal projection of the tilted part of the structure. Therefore, the structure must also withstand large deformations and at the same time exhibit the greatest possible auxetic behavior. In this way the geometry of the structure also has an influence on how much the structure is elongated, and the Poisson's ratio is evaluated always after the above $\Delta L_x$ is reached.	

In the present example, a different optimization strategy is used than in the Examples \ref{ssec_Ex1} \& \ref{ssec_Ex2}. In the previous examples, it was found that the Nelder-Mead optimization method seems to provide the best estimate of the optimum, albeit not in a consistent and robust manner. That is, several optimization runs were required to find the optimum. Therefore, we choose the Nelder-Mead method as the only optimization method here, and the optimization is repeated 20 times. As before, in the case when the strain in a truss element is outside the training domain of the NN, i.e. if the stress constraint is violated, the analysis fails due to poor quality of NN prediction. In this case, the Poisson's ratio was assigned a small positive number ($10^{-6}$), while the corresponding stress constraint are handled in the same way as in Ex.~\ref{ssec_Ex2} to penalize such a solution.

The preliminary optimization runs have shown that the fitness term which includes the Poisson's ratio is dominant. Therefore, the weights are chosen as described above, i.e. with reduced influence of the Poisson's ratio term. It also became clear that due to Eq.(\ref{eq_Ex3_Poisson}) structures with larger $L_\mathrm{x}$ will be more favorable. The remaining questions concerned the influence of the allowable stresses and the total length of the rods on the feasibility of the solutions. In particular, it was noticed that the lower values of $L_\mathrm{mid}$ near both optima lead to the violation of the compressive stress constraint. The last constraint was quite low due to the large CNT diameter and therefore quite sensitive to the shell-like buckling.

Although the influence of the Poisson's term on the fitness function has already been reduced, it still remains quite dominant. In particular, decreasing $L_\mathrm{hor}$, $L_\mathrm{mid}$ will reduce $L_\mathrm{tot}$, but the reduction of the Poisson's term when they increase will be more important. For this reason, the global optimum obtained by multiobjective optimization corresponds to the optimum obtained by minimizing the Possion's ratio in the single objective optimization. The local optimum then represents a compromise between the minimum Poisson's ratio and the minimum total length. At the global optimum it turns out that $L_\mathrm{hor}$, $L_\mathrm{mid}$ should be at the upper bounds, while $h$ and $\alpha$ should be at the lower bounds. The remaining parameter $t$ at the global optimum is around $4.8$ nm. 

\begin{figure}
	\centering
	\includegraphics[scale=0.10]{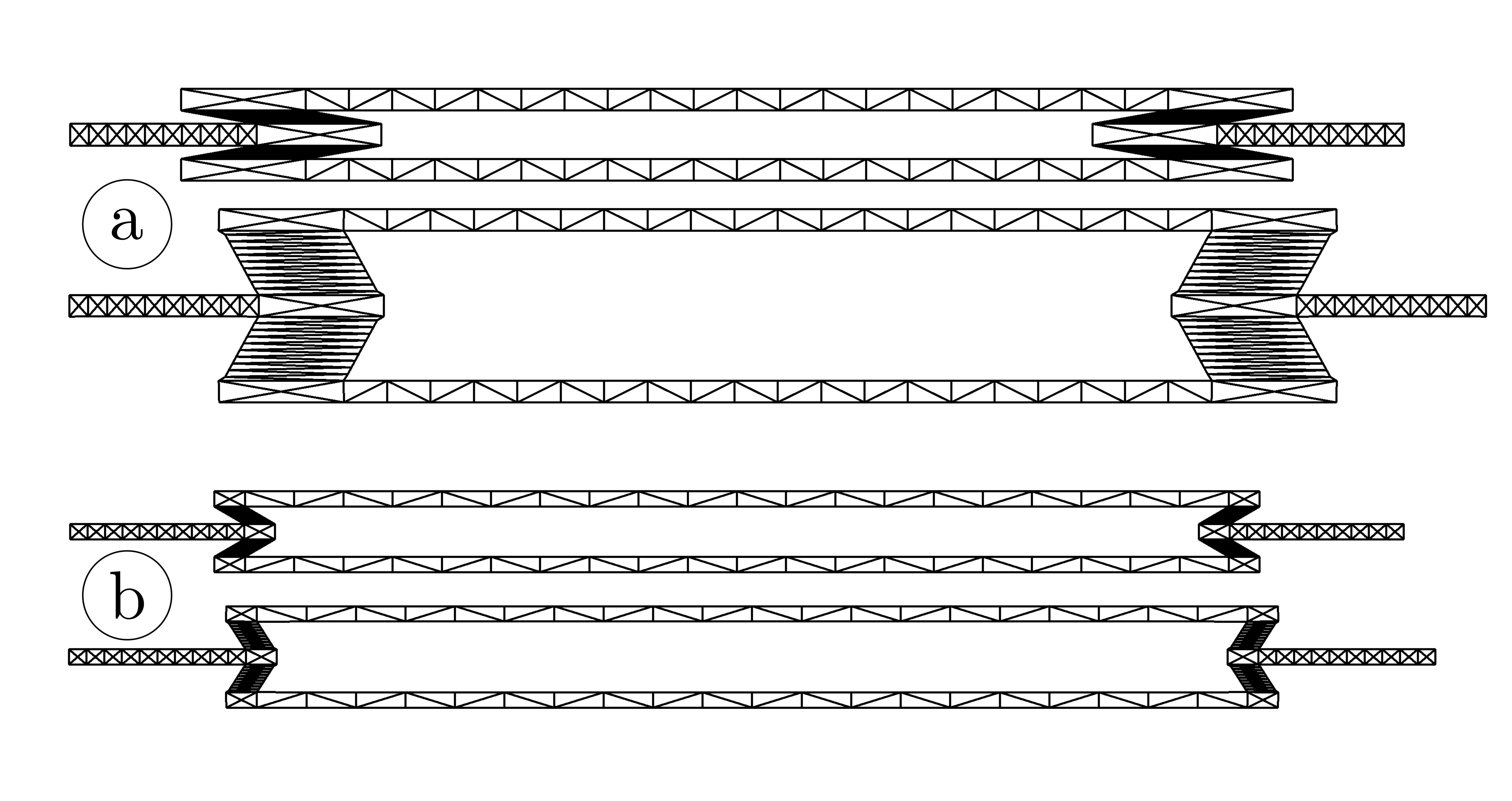}
	\caption{Initial and final configuration: (a) global optimum and (b) local optimum.} 
	\label{fig_ex3_globlocconf}
\end{figure}

\begin{figure}
	\centering
	\includegraphics[scale=0.40]{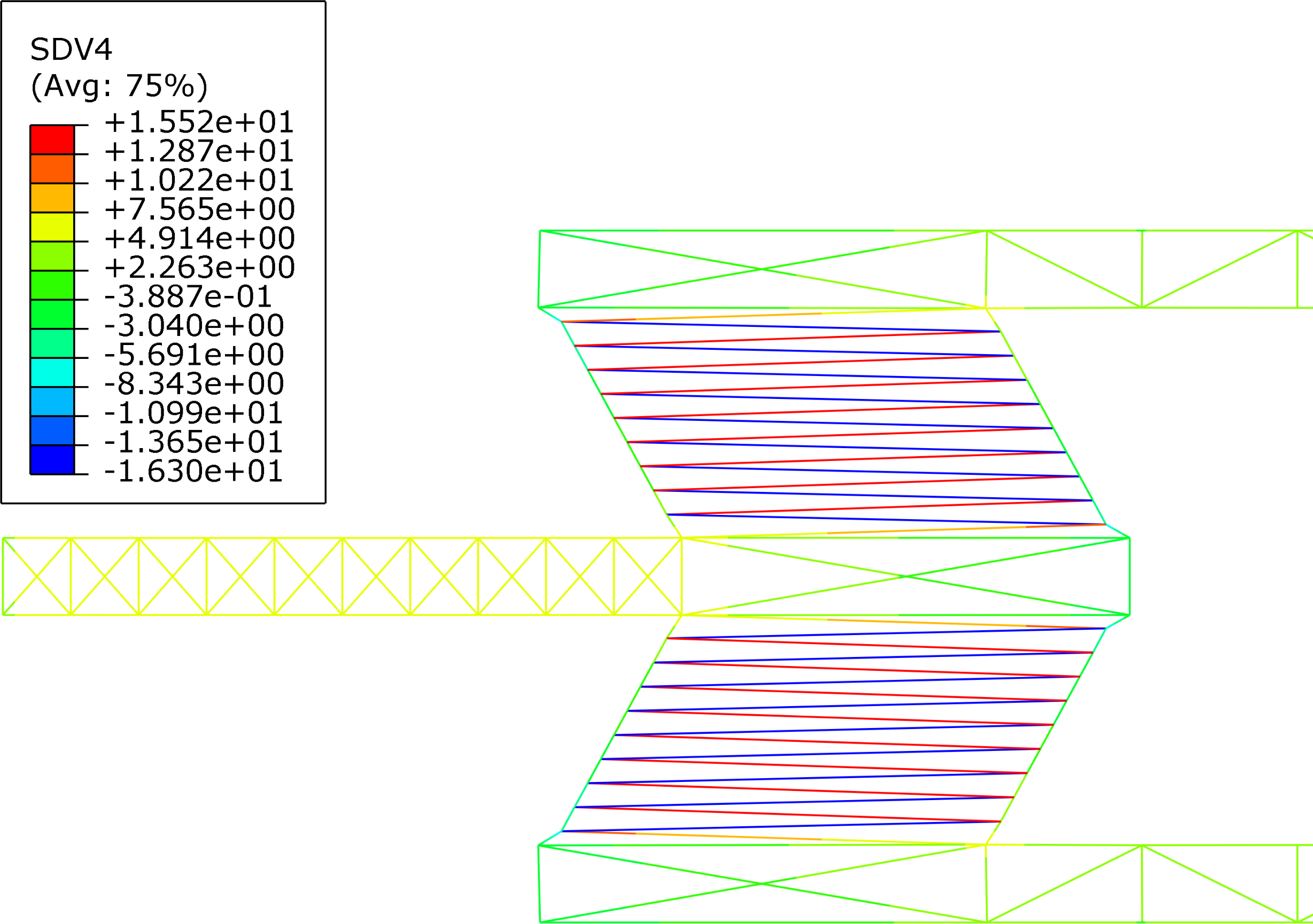}
	\caption{Global optimum - a detail of the final configuration with the true stress distribution (GPa).} 
	\label{fig_ex3_globlocconf_detail}
\end{figure}

The optimization procedure clearly converged to two optima: a global and a local one, Tab.~\ref{tab_Ex3_optim} and Fig.~\ref{fig_ex3_globlocconf}. An extreme Poisson's ratio of approximately -17.8 was achieved for the global optimum. A detail of the final global optimum configuration, Fig.~\ref{fig_ex3_globlocconf_detail}, shows that the critical part of the nanostructure are the rods in the ribs, which are loaded in compression. In the case of the local optimum, the axial stresses in these rods are equal to the minimum permissible compressive stresses. The local optimum corresponds to a Poisson's ratio of approximately -10.5, and the total length was also reduced by more than double.

The most important statistical parameters are summarized in Tab.~\ref{tab_Ex3_stat}. A very low value of the standard deviation indicates that the optimization procedure was quite consistent and robust for both minima. The probability of finding a global optimum was clearly higher. For the reader's convenience, a correlation analysis is provided in \ref{sec:app3}. Note that three analyzes yielded optima with $\alpha\approx10\circ$, but other variables were far from the global optimum and were therefore disregarded in the statistical analysis.

\begin{figure}
	\centering
	\includegraphics[scale=0.5]{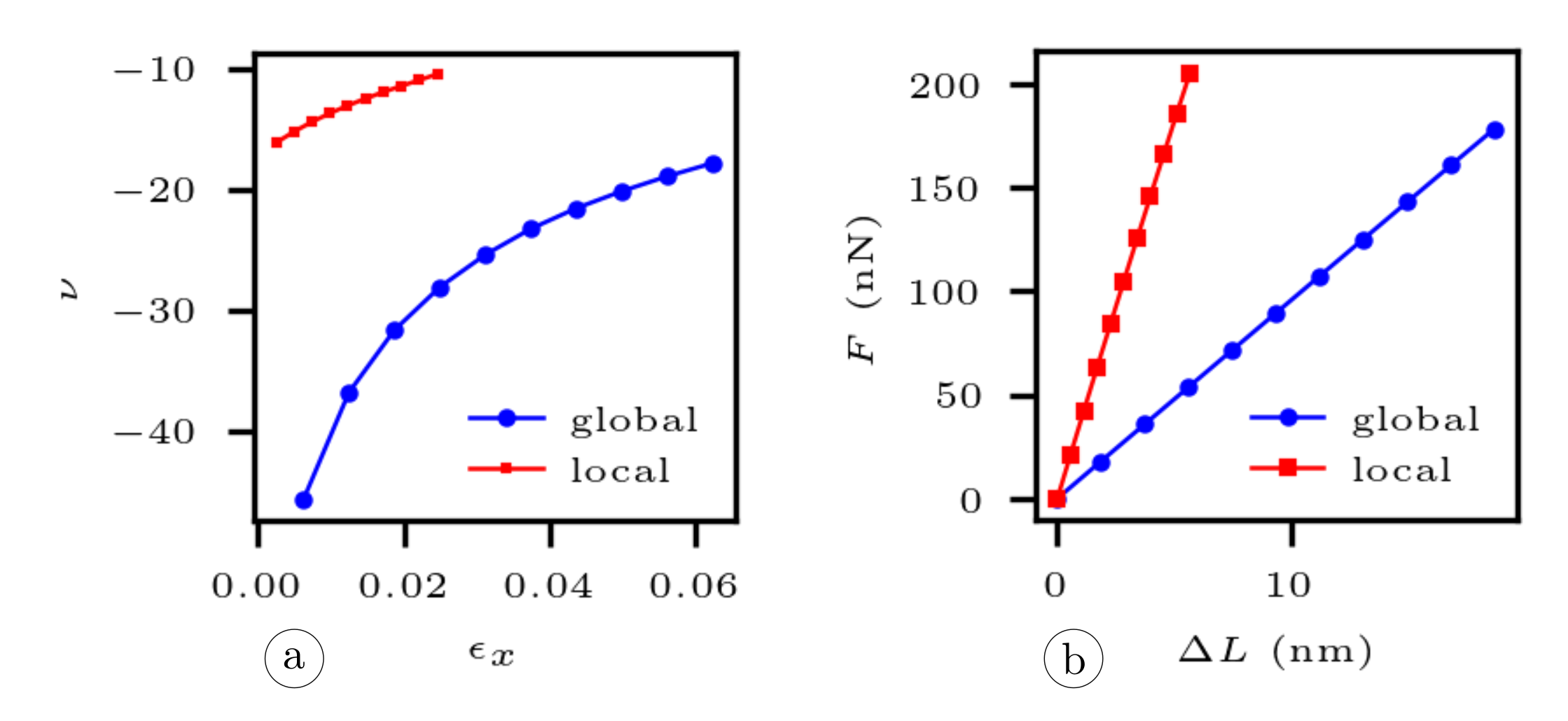}
	\caption{(a) Dependence of the Poisson's ratio on the longitudinal strain for the global and local optimum; (b) Force - elongation curves for the global and local optimum.} 
	\label{fig_ex3_5_FDL}
\end{figure}

It has already been mentioned above that the Poisson's ratio is not constant during deformation. This dependence is shown in Fig.~\ref{fig_ex3_5_FDL}a. As expected, the Poisson's ratio is highest at the beginning and lowest at the end. It ranges from $\nu=-45.67$ to $\nu=-17.82$ for the global optimum and from $\nu=-16.16$ to $\nu=-10.46$ for the local optimum. At the end, the force-elongation curves for both optimal configurations are shown in Fig.~\ref{fig_ex3_5_FDL}b. The load is applied in the horizontal direction. An almost linear response was observed, with the local optimal configuration having a higher stiffness in the horizontal direction.

\begin{table}
	\begin{center}
		\addtolength{\leftskip} {-2cm}
		\addtolength{\rightskip}{-2cm}
		{		\footnotesize
			\begin{tabular}{c|c|c|c|c|c|c|c|c|c|c}
				\hhline{=|=|=|=|=|=|=|=|=|=|=}
				$h$ & $t$ & $\alpha$ & $L_\mathrm{hor}$ & $L_\mathrm{mid}$ & $\nu$ & $L_\mathrm{tot}$ & fit. & $\sigma_\mathrm{min}$ & $\sigma_\mathrm{max}$&  eval. \\
				(nm) & (nm)  &($\circ$) &  (nm)  & (nm) &  & (nm)  & & (GPa) &  (GPa) &    \\
				\hline
				Global & &   &   &  & & & &  &   &  \\
				3.0 &	4.868 & 10.003	& 25.0 & 79.999 & -17.82 &	4853.8 & 0.122  & -16.30 & 15.52 & 1001 \\
				Local &   &   &   &  & & & &  &   &  \\
				3.0 &	2.660 & 30.166	& 25.0 & 80.000 & -10.46 &	2070.1 & 0.163  & -19.53 & 17.61 & 849 \\
				\hline 				
				Bounds &   &   &   &  & & & &  &   &  \\
				3	& 2.656 & 10 & 2 &	5 & & & & & & \\
				60	& 22.163 & 80 & 25 & 80 & & & & & & \\
				\hhline{=|=|=|=|=|=|=|=|=|=|=}
			\end{tabular}
		}	
		\caption{The best global minimum, the best local minimum, and upper and lower bounds on variables.}
		\label{tab_Ex3_optim}
	\end{center}
\end{table}

\begin{table}
	\begin{center}
		{		\footnotesize
			\begin{tabular}{c|c|c|c||c|c|c}
				\hhline{|=|=|=|=|=|=|=|}
				& Global &  &   &  Local &  &  \\
				&  $\nu$ & $L_\mathrm{tot}$  & Fit & $\nu$ & $L_\mathrm{tot}$  & Fit \\
				\hline
				median	& -17.764 &	4853.8 & 0.123 & -10.460 & 2070.1 & 0.163  \\
				average	& -17.732 &	4832.7 & 0.123 & -10.452 & 2074.2 & 0.163  \\ 
				st. dev.&   0.101 &	43.9 &	$1.46 \cdot 10^{-4}$ & 0.022 &	11.61 &	$7.18 \cdot 10^{-4}$ \\ 	
				\hline
				Number of minima & 13 & &  & 4  & &\\ 			
				\hline			
				Evaluations & & & & & & \\
				median  & 1001 & & & 849 & & \\			
				average & 923 & & & 879 & & \\				
				\hhline{|=|=|=|=|=|=|=|}
			\end{tabular}
		}	
		\caption{Statistical data on global and local minima.}
		\label{tab_Ex3_stat}
	\end{center}
\end{table}

\subsection{Shape Optimization of a Nanotruss Cantilever}
\label{ssec_Ex4}
\begin{figure}
	\centering
	\includegraphics[scale=0.29]{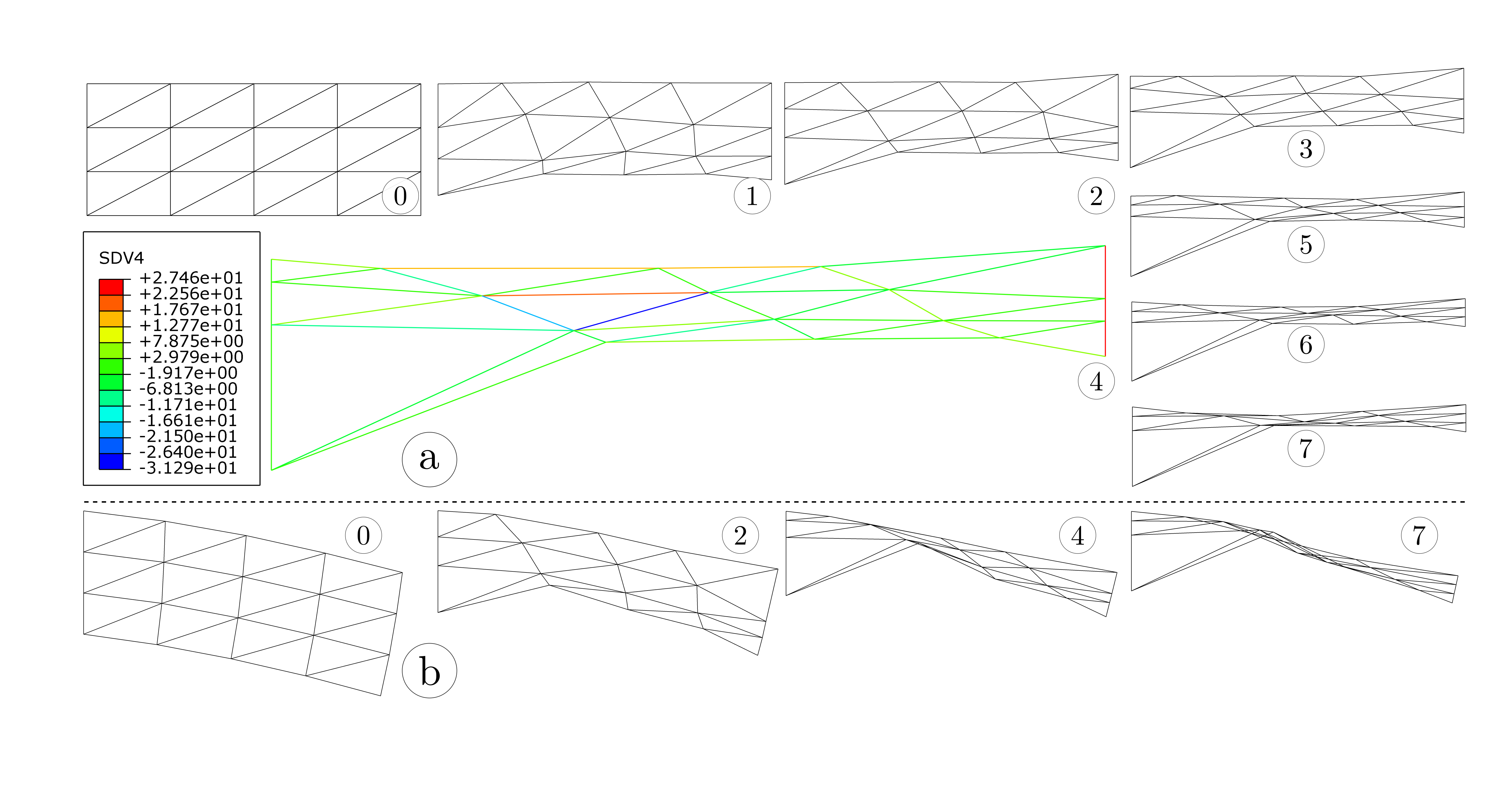}
	\caption{Configurations: (a) undeformed at each optimization step; at step (4) the axial stresses in the truss are shown; (b) selected deformed configurations.} 
	\label{fig_ex4_configs}
\end{figure}

The example at hand considers a computationally much more complex problem, a shape optimization of a cantilever truss, Fig.~\ref{fig_ex4_configs} (step 0). Optimizing this nanotruss requires a slightly different approach, and that is the reason for its consideration in this research. The initial truss is $L=50$ nm long and $h=20$ nm high and consists of (10,10) SWCNTs. The initial diameter of this nanotube at 300 K is $D_0=1.3374$ nm. As usual, the thickness of the nanotube is $t=0.34$ nm. The minimum compressive stress is $\sigma_\mathrm{c}=-37.59$ GPa, while the maximum tensile stress is $\sigma_\mathrm{t}= 104.61$ GPa.

The nanotruss is assembled out of $N_\mathrm{e}=43$ truss elements and 40 nodes. The leftmost nodes of the cantilever are fixed, while at the right end downward displacement of 10 nm is applied.

The optimization problem is defined as follows:
\begin{equation}
	\label{eq_Ex4_opt_prob}
	\begin{array}{llc}
		\underset{\mathbf{x}}{\text{minimize}} &  L_\mathrm{tot}(\mathbf{x}) &\\
		\text{subject to} &   \sigma_i \le\sigma_\mathrm{t} & (\text{a}) \\
		&    \sigma_i \ge \sigma_\mathrm{c} & (\text{b}) \\
		&   L_{i} \ge D_0 & (\text{c})\\
		&   L_{i} \le 12.5 D_0, \quad \text{if} \quad \sigma_i\le 0 & (\text{d}). \\
	\end{array}	
\end{equation} 
where $i=1,..N_\mathrm{e}$. Thus, the objective is to reduce the total length of the rod elements $L_\mathrm{tot}$ by varying nodal coordinates, providing a new truss shape. In this way, the set of variables $\mathbf{x}$ is composed of the nodal coordinates. The only exceptions are the abscissas of the nodes where supports or displacements are applied which are not part of the variables set. Note that the ordinates of these nodes can vary and are included in the latter set. In order to fix the nanotruss is space, the coordinates of the node in the lower left corner are also excluded from the optimization process. This gives the total of 31 variables. As far as the constraints are concerned, these are now placed on the stresses at the end of simulation and on the lengths of the individual elements. The total number of constraints is 171.

This example can be classified as a medium sized optimization problem and requires a more elaborate approach. The pseudo-code of the procedure is given in Algo.~\ref{Ex4_algo}. The cornerstone is the Nelder-Mead optimization method. 

In the first step, two successive runs are performed. Before each run, a series of random candidates is evaluated and the best results are used as the initial variables. After each run, the finite element mesh is modified by adding variables to the corresponding node coordinates. This serves as the base configuration for the next optimization run. Note that this also involves redefining the bounds imposed on the variables and is not the same as updating variables while maintaining the fixed bounds. This series of two runs is performed 10 times and the best result is selected as the basic configuration for the next step. The procedure is then continued in five runs, updating the base configuration between each run. Each of these 5 runs is repeated three times, again selecting the best result as the next base configuration. As the optimization progresses, the limits within which the variables can vary and the number of allowed evaluations are reduced. The number of stalled evaluations that terminate the procedure has been set at 300.

Such an optimization procedure requires further clarification. The present example involves both material and geometric nonlinearities. Since large displacements are enforced, a highly irregular configuration in most cases will lead to failed convergence and an incomplete analysis. These results are unusable for the optimization. For example, the first run was preceded by evaluation of 100 random candidates, of which about two thirds resulted in failed analyzes. Since increasing the lower and upper limits for the variables contributes to an increased irregularity of the structure, setting these bounds too high would essentially prevent the optimization from finding the solution. On the other hand, the initial configuration is expected to be changed significantly to obtain the minimum total length. For this reason, the initial bounds for the variables are somewhat lower than the actual final bounds, and the optimization process can then converge more smoothly to an intermediate or local minimum. In the second and subsequent steps, the base configuration is then updated and new, lower limits for the variables are applied to the new configuration. Although lower, these limits are still high enough to detect another lower minimum if the first step has converged to a local minimum. Such a procedure allows for variable changes that would otherwise be too large for a single step.

When it comes to interpreting the results, the difference between the problem of prescribed displacement and that of prescribed force should first be emphasized. From the optimization point of view, these two cases are fundamentally different. Both types of boundary conditions will lead toward reduction in the stiffness of the nanotruss structure. In the force-driven problem, the forces must remain balanced in each node as the optimization progresses, and since the external forces are constant, the stress constraints become increasingly important. This limits the number of feasible solutions. On the other hand, the reduction of stiffness in the displacement-driven problem leads to a reduction of reaction forces at the nodes where the displacements are prescribed. As a result, the stresses in the structure are also reduced and consequently the stress constraints are relaxed. Thus, new feasible solutions can emerge and the optimization process can now converge to more local minima which can be a significant complication. Eventually, the displacement-driven problem will provide a very thin and flexible structure, with a minimal mass and low stresses, Fig.~\ref{fig_ex4_configs}a (step 7).

The initial (step 0) and the nanotruss configuration obtained after each step are shown in Fig.~\ref{fig_ex4_configs}a, while selected deformation shapes are shown in Fig.~\ref{fig_ex4_configs}b. Note that in steps 4-7 rods cross each other, which would require the implementation of some kind of contact algorithm. Note that this would also involve bending of the nanotubes, which cannot be simulated with the truss elements used here. For this reason, configurations in which such behavior is barely present could be the chosen as the design configuration. Nevertheless, the trend of decreasing stiffness is clearly visible ending in a very thin structure at the right end of the nanotruss.

\begin{figure}
	\centering
	\includegraphics[scale=1.4]{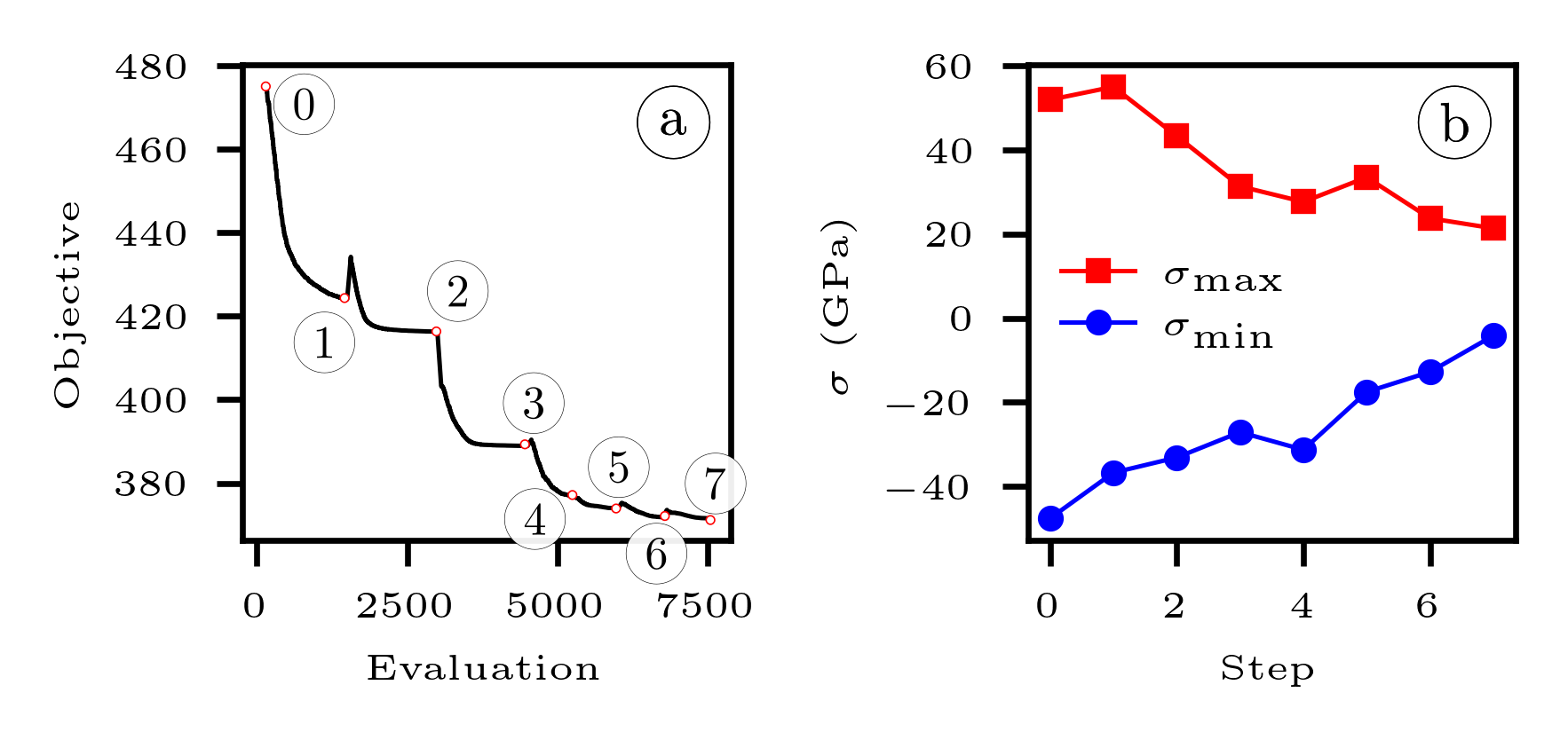}
	\caption{(a) Objective vs. evaluations. The numbers indicate the configurations in each optimization step, Fig.~\ref{fig_ex4_configs}; (b) maximum and minimum stresses in each step.} 
	\label{fig_ex4_diags}
\end{figure}

The convergence properties of such a procedure can be seen in Fig.~\ref{fig_ex4_diags}a. It is obvious that each step tends to converge to a certain value of the objective. At the beginning of a new step, the new objective is somewhat worse than the old one, which is attributed to the initial evaluations of the random candidates. With the exception of the last step, all steps ended with the maximum number of allowed evaluations. In the last step, the optimization was stopped in 2 out of 3 runs due to 300 stalled evaluations, which is an indication that further progress is only possible with further fine-tuning of the optimization parameters. As expected, the maximum and minimum stresses in the rods are reduced in each step, Fig.~\ref{fig_ex4_diags}b, steadily increasing the importance of the length constraints compared to the stress constraints. As for the distribution of stresses, Fig.~\ref{fig_ex4_configs}b step 4, these are fairly evenly distributed in the nanotruss.

\newcommand{\forcond}{$i=0$ \KwTo $n$}
\SetKwFunction{FRecurs}{FnRecursive}%
\SetStartEndCondition{ }{}{}%
\SetKwProg{Fn}{def}{\string:}{}
\SetKwFunction{Range}{range}
\SetKw{KwTo}{in}
\SetKwFor{For}{for}{\string:}{}%
\SetKwIF{If}{ElseIf}{Else}{if}{:}{elif}{else:}{}%
\SetKwFor{While}{while}{:}{fintq}%
\renewcommand{\forcond}{$i$ \KwTo\Range{$n$}}
\AlgoDontDisplayBlockMarkers
\SetAlgoNoEnd
\SetAlgoNoLine
\SetKw{Len}{len}

\SetKw{Return}{return}
\SetKwComment{Comment}{\#}{}
\SetKwBlock{Begin}{}{}
\SetKwProg{Function}{function}{}{}

\IncMargin{1em}
\begin{algorithm}
	\DontPrintSemicolon
	${N}_\mathrm{steps} \longleftarrow 7$\;   
	$\mathbf{N}_\mathrm{run} \longleftarrow \left\lbrace 10, 3, 3, 3, 3, 3 \right\rbrace $\;
	$\mathbf{N}_\mathrm{eval} \longleftarrow \left\lbrace 1500, 1500, 1500, 750, 750, 750, 750 \right\rbrace $\;
	$\mathbf{N}_\mathrm{ini} \longleftarrow \left\lbrace 100, 100, 30, 30, 30, 30, 30 \right\rbrace $\;
	$\boldsymbol{\omega} \longleftarrow \left\lbrace 1, 1, 0.5, 0.5, 0.25, 0.25, 0.20 \right\rbrace $\;
	Define base variable bounds $\Delta \mathbf{x}_0=3.33 \mathbf{I}$\;
	\Function{optimization\_step(i)}{
		Define upper and lower bounds for variables as $\mathbf{x}_\mathrm{u}=\omega_i \Delta \mathbf{x}_0$, $\mathbf{x}_\mathrm{l}=-\omega_i \Delta \mathbf{x}_0$\;
		Run $N_{\mathrm{ini},i}$ initial FE simulations and take the best configuration as an initial guess for variables $\mathbf{x}_i$\;
		Run optimization for $N_{\mathrm{eval},i}$ evaluations to obtain optimal variables $\mathbf{x}_i$ for the current step $i$\;
		Update initial nodal coordinates $\mathbf{X}_0 \leftarrow \mathbf{X}_0+\mathbf{x}_i$\;
		\Return $\mathbf{X}_0$
	}\;
	\For{$i$ \KwTo \Range$(N_{\mathrm{run},0})$}{
		Set nodal coordinates $\mathbf{X}_0$ to initial values\;
		$\mathbf{X}_0$=optimization\_step(0)\;
		$\mathbf{X}_0$=optimization\_step(1)\;
		Store nodal coordinates from step $i$\;
	}
	Use the best nodal coordinates $\mathbf{X}_0$ as the new coordinates\;
	\For{$i$ \KwTo \Range$(2,N_\mathrm{steps})$}{
		\For{$j$ \KwTo \Range$(N_{\mathrm{run},i-1})$}{
			$\mathbf{X}_0$=optimization\_step($i$)\;
			Store nodal coordinates from step $j$\;
		}
		Use the best initial nodal coordinates $\mathbf{X}_0$ as the new coordinates\;
	}
	\caption{The shape optimization algorithm used in Ex.~\ref{ssec_Ex4}; {\footnotesize  ${N}_\mathrm{steps}$ - number of optimization steps, $\mathbf{N}_\mathrm{run}$ - number of optimization runs per step, $\mathbf{N}_\mathrm{eval}$ - number of evaluations in each run of a step, $\mathbf{N}_\mathrm{ini}$ - number of initial random candidates, $\boldsymbol{\omega} $ - weighting factors for variable bounds.} \label{Ex4_algo} }
\end{algorithm}
\DecMargin{1em} 

\subsection{Increasing the compressive strength of a CNT nanotruss based metamaterial}
\label{ssec_Ex5}
\begin{figure}
	\centering
	\includegraphics[scale=0.14]{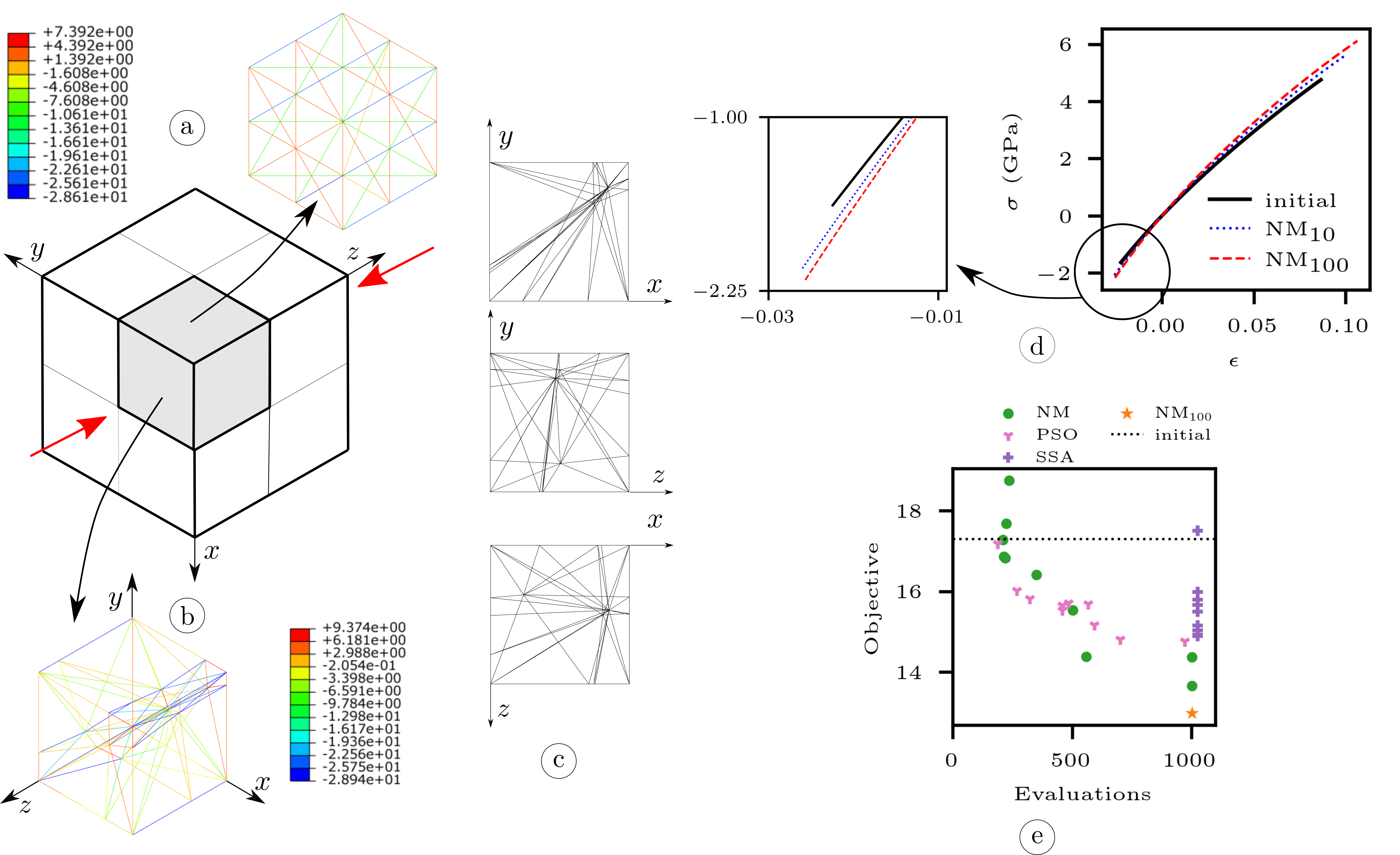}
	\caption{Compression of an RVE. (a) A cube representing a microstructure is divided into octets and on selected octet RVE periodic and symmetry boundary conditions are applied. The colors indicate the true axial stresses in the truss elements in an initial (non-optimal) RVE; (b) the best NM achieved in 100 optimizations (NM$_{100}$); (c) different views of the NM$_{100}$ configuration; (d) tensile and compressive engineering stress-strain curves of the RVE for the initial configuration and the best NM configurations out of 10 (NM$_{10}$) and 100 runs; (e) objectives and number of evaluations for the considered optimization methods; for NM only first 10 results and the global optimum NM$_{100}$ are shown.} 
	\label{fig_ex5}
\end{figure}

The last example deals with the design of an orthotropic metamaterial in which a basic repeating structure is a three-dimensional carbon nanotube nanotruss. The main idea is to demonstrate the possibility of handling more complex materials than those with simple structures such as SC or FCC, which have already been analyzed in the literature, see for example \cite{zhang2018nano}. Since CNT are particularly sensitive to buckling, the compressive strength of a CNT is more critical than the tensile strength. In the present case, the compressive strength of the initial representative volume element (RVE), Fig.~\ref{fig_ex5}a, should be increased as much as possible in the $z$ direction. 

The objective function should be defined first. To this end, note that FE simulations do not fail during optimization exactly when the compressive strength of a rod is reached. The NN still manages to provide stresses even after the compression stress limit has been exceeded, but with reduced accuracy. For this reason, the exact compressive strength for the configuration at hand should be somehow interpolated, which raises new questions about how this should be done during the optimization. We circumvent this by using an indirect approach. The main objective here is to reduce the compressive stress in the critical element as much as possible. This is still a demanding optimization process since the critical element is not known in advance and during the optimization the particular critical element is often changed, leading to a discontinuity in the solution process. One last remark: the fracture of a single rod does not necessarily mean that the whole RVE will fail. The internal forces could be redistributed in the remaining rods. The point at which the first rod fails could therefore be understood as the counterpart of a yield point in metals.

The RVE under consideration is a cube with a side length of 40 nm, Fig.~\ref{fig_ex5}a. Periodic boundary conditions are used. Furthermore, orthotrophy is assumed, which allows the application of the symmetry boundary condition to faces parallel to the $xz$ and $yz$ planes and the decomposition of the cube into octets. Only one octet is of interest, and the edge of the octet under consideration is 20 nm long. The force $F=400$ nN was applied in the direction of the $z$-axis, at the vertex with coordinates (0,0,0). (15,15) SWCNTs were selected as truss elements, with properties $D_0=2.00235$ nm, $A=2.1377$ nm$^2$, $\sigma_\mathrm{c}=-28.52$ GPa, $\sigma_\mathrm{t}=132.06$ GPa. The length delimiting shell-like and beam-like buckling is $L_\mathrm{max}=12.5D_0= 25.0294$ nm.

The optimization problem is to keep the absolute value of the minimum stress in a critical element as low as possible. Consequently, the objective function and the associated constraints are as follows:
\begin{equation}
	\label{eq_Ex5_opt_prob}
	\begin{array}{llc}
		\underset{\mathbf{x}}{\text{minimize}} &  \left| \min {\sigma_i}\right|  (\mathbf{x}) &\\
		\text{subject to} &   \sigma_i \le\sigma_\mathrm{t} & (\text{a}) \\
		&    \sigma_i \ge \sigma_\mathrm{c} & (\text{b}) \\
		&   L_{i} \ge D_0 & (\text{c})\\
		&   L_{i} \le 12.5 D_0, \quad \text{if} \quad \sigma_i \le 0 & (\text{d}). \\
	\end{array}	
\end{equation} 
where $i=1,..N_\mathrm{e}$, $N_\mathrm{e}=99$ is the total number of rods in the nanotruss. The number of nodes is 27. In this way, there are 395 constraints that have the same form as in Ex.~\ref{ssec_Ex4}. The objective is to be achieved by shifting positions of selected nodes $\mathbf{x}$. The vertices of the cube are fixed in space, the nodes on the edges can only move on the edge, while the nodes on the faces can only move in the corresponding plane of this face. The node within the cube can move freely, but must remain within the cube. The limits for all nodes ensure that all nodes remain within the original dimensions of the cubic RVE. This results in a total of 27 variables. 

The analyses regarding to choice of the suitable optimization methods presented in Exs.~\ref{ssec_Ex1}, \ref{ssec_Ex2} are revisited in the present example. The optimization methods in question were NM, PSO and SSA, with the swarm/population size corresponding to the number of variables. Each optimization was run 10 times, with the maximum number of evaluations of \numprint{1000} and the maximum number of stalled evaluation 100. A total of 30 optimization runs were therefore performed. The performance of the optimization method is shown in Fig.~\ref{fig_ex5}e. It is clear that SSA shows a very robust and consistent performance. The realized objectives are closely packed, with the least deviation in the results. On the other hand, SSA does not seem to be affected by the stalled optimization, so a larger number of evaluations could lead to a better solution. PSO seems to be affected by the stalled evaluations, so a larger allowed number of stalled evaluations could be beneficial. PSO and NM show a larger dispersion of results, indicating less consistency. This is particularly true for NM, which was also observed in other examples. However, the wider set of solutions could be more advantageous, especially for a discontinuous problem like this one. In particular, the best three out of 30 sets of variables are obtained by NM (and the worst two as well). This shows once again that the best approach is to run the NM method for a selected number of runs to obtain the variables that yield the lowest objective. For this reason, NM was run for another 90 runs, and the best result can be seen in Fig.~\ref{fig_ex5}b. At first glance, the configuration appears very irregular, but if different views are considered, Fig.~\ref{fig_ex5}c, the interpretation is easier. It seems that the main feature of the new configuration is that the CNT rods are shifted to an edge parallel to the $z$ axis, more precisely to the center of the RVE from which the octet is extracted. The shortest CNT length constraint does not allow any further repositioning of the CNT towards the particular edge of the octet.

\npdecimalsign{.} 
Standard tensile and compression tests were performed on the RVE for the initial configuration and for the best results of the first 10 (NM$_{10}$) and 100 NM optimization runs (NM$_{100}$), Fig.~\ref{fig_ex5}b. The engineering stress-strain curves obtained are shown in Fig.~\ref{fig_ex5}d. The area used for the stress calculation was the initial area of the RVE face, while the initial edge length was used to calculate the strains. Although the aim was to increase the compressive strength compared to the initial configuration, which is clearly achieved (by 21.5\% for NM$_{10}$ and by 25.6\% for NM$_{100}$), the tensile strength is also increased (by 14.0\% for NM$_{10}$ and by 21.1\% for NM$_{100}$). In both tensile and compressive regimes, RVE always fail due to buckling of a rod. In other words, even in the tensile regime, there will always be a rod that is critically loaded in compression. The stiffness is also slightly increased.  

The properties of the present design can be now compared to steel. There are about 118 atoms/nm in (15,15) SWCNT, which gives a specific mass of $2.3553\cdot 10^{-24}$ kg/nm. Then the length of the NM$_{100}$ configuration was \numprint{1282.1} nm (the equivalent mass is $3.020\cdot 10^{-21}$ kg). This results in a specific mass density of $377.5$ kg/m$^3$. Finally, the tensile ($\sigma_\mathrm{t}=6.11$ GPa) and compressive ($\sigma_\mathrm{c}=-2.17$ GPa) strength expressed as specific quantities are then 16.152 MPa/kg and -5.756 MPa/kg respectively. For comparison, the specific tensile strength of AISI 304 stainless steel is approximately 0.063 MPa/kg. Although this is not a fair comparison, as the CNT-based nanotruss considered here is defect-free, whereas the data for AISI 304 comes from experimental data involving a crystal structure with defects, this difference is quite extreme by a factor of 232. 

The initial Young's moduli is calculated from the stress-strain curve and is 74.9 GPa for NM$_{100}$. This results in a ratio between tensile strength and Young's modulus as $\sigma_\mathrm{t}/E=0.082$, which is much higher than for metals \cite{ashby1999material}.

\section{Conclusion}
The main contribution of the research at hand is the application of optimization techniques and methodologies for the design of new metamaterials and nanoscale structures at room temperature. Carbon nanotubes are the central element of the procedure. CNT are assembled into a nanotruss structure, combining the unique mechanical properties of CNT with the exceptional loading capacity of trusses. Such a procedure builds on a recently developed MD-NN-FEM framework \cite{canadija2024computational}. This framework is reliable and computationally efficient, enabling repeated simulations within a reasonable time frame, which is required in any optimization procedure.

Several popular optimization methods have now been tested in conjunction with the above framework: ABC, BA, NM, PSO and SSA. It turned out that ABC and BA did not perform well on the problems at hand. PSO and SSA performed well, although a larger number of evaluations or allowed stalled evaluations could be helpful. However, NM \cite{nelder1965simplex} was the only optimization method that could provide the best optima in most cases, but with much less robustness than PSO or SSA and with much larger deviations. Due to the greater variability of the solutions, a larger number of runs is therefore required for NMs.

In the Examples section, we have shown that such an optimization framework can indeed yield CNT-based nanotruess with superior performance. Apart from the introductory example that is intended to determine the most appropriate optimization method for the task, three examples focused on the development of new metamaterials, while one example demonstrated the possibility of structural design at the nanoscale. It is shown that a CNT-based metamaterial can be designed to trap significant energy or to exhibit extraordinary auxetic behavior with a Poisson's ratio as low as -17.82. Similarly, with the compressive strength of CNT being the critical issue, it is shown that an anisotropic metamaterial with a compressive strength in one direction can achieve the theoretical strength of 5.756 MPa/kg.

Finally, there are few caveats to point out. Firstly, nanotechnology can still only produce the simplest CNT nanotrusses \cite{portela2021supersonic}. Nevertheless, we believe that the present results motivate further research on the fabrication of more complex structures. Secondly, all investigations in this work assume ideal CNT \cite{canadija2021deep}, without any kind of defects. This is certainly not the case in the real world. Taking defects into account will reduce the present results by a certain factor, but should still give a significant difference to conventional materials. Third is the approximation that CNT joints can be treated as hinges, but the alternative approximation that would require beam finite elements, i.e. a rigid joint, is also a simplification. It remains to be seen what is closer to reality. The improvement of the model to better account for this issue would be based on beam elements with flexible connections. Lastly, the present simulations are quite computationally intensive. For this reason, we have avoided using a large number of optimization runs or a larger number of evaluations. Increasing these would certainly lead to better solutions.

\section*{Acknowledgments}
This work has been supported in part by the Croatian Science Foundation under the project IP-2019-04-4703 and in part by the University of Rijeka under project
number uniri-iskusni-tehnic-23-37. This support is gratefully acknowledged. 

\section*{CRediT author statement}
\textbf{Marko \v{C}ana\dj{}ija:} Conceptualization, Funding acquisition, Investigation, Methodology, Project administration, Software, Validation, Writing - Original Draft, Writing - Review \& Editing. \textbf{Stefan Ivi\'{c}:} Investigation, Methodology, Software, Validation, Writing - Original Draft, Writing - Review \& Editing.

\appendix
\section*{Appendices}

\section{Addendum to Example 1}
\label{sec:app1}

The Tab.~\ref{tab_Ex1_optim} summarizes several optimization results of interest discussed in the main part of the text.
\begin{table}
	\begin{center}
		\addtolength{\leftskip} {-2cm}
		\addtolength{\rightskip}{-2cm}
		{		\footnotesize
			\begin{tabular}{c|r|r|r|c|r|r|r|r|r|r|r}
				\hhline{=|=|=|=|=|=|=|=|=|=|=|=}
				opt. & $x_A$ & $y_A$ & $y_B$  & obj. &$\sigma_1$ & $\sigma_2$& $\sigma_3$ & $L_1$ & $L_2$ &$L_3$ & eval.\\
				\hline
				NM$^1$     & 0.031 & -0.782 & -1.262 & 12.902 &  84.8 &  39.3 &  84.8 & 0.78 & 5.99 & 6.13 & 892 \\ 
				NM$^2$     & 0.031 & -0.782 & -1.262 & 12.902 &  84.7 &  39.3 &  84.8 & 0.78 & 5.99 & 6.13 & 1002 \\ 
				SSA$^3$    & 0.106 & -0.888 & -1.378 & 12.965 &  84.5 &  39.2 &  84.6 & 0.89 & 5.91 & 6.16 & 1010 \\
				PSO$^4$    & 0.310 & -1.107 & -1.518 & 13.044 &  84.7 &  40.2 &  84.2 & 1.15 & 5.70 & 6.19 & 930 \\
				BA$^5$     & 0.088 & -0.902 & -2.154 & 13.324 &  84.2 &  37.6 &  79.5 & 0.91 & 6.04 & 6.37 & 670 \\
				NM$^6$     & 0.959 &  2.903 &  6.892 & 18.623 & -69.7 & -17.8 & -69.7 & 3.06 & 6.43 & 9.14 & 863 \\
				ABC$^7$    & 1.496 & -2.918 & -4.389 & 15.452 &  84.2 &  37.5 &  66.7 & 3.28 & 4.74 & 7.43 & 455 \\
				PSO$^{8}$    & 1.248 &  4.469 &  7.438 & 19.800 & -69.7 & -19.6 & -67.6 & 4.64 & 5.60 & 9.56 & 370 \\
				ABC$^{9}$    &-2.561 &  1.291 & -7.888 & 26.570 & \textit{-101.7} &  49.4 &  78.7 & 7.73 & 12.55 & 6.29 & 375 \\
				SSA$^{10}$ &-3.011 &  9.309 &  7.654 & 28.671 & -67.8 &  24.8 & \textit{-72.3} & 9.78 & 9.16 & 9.73 & 430 \\
				NM$^{11}$  &-3.057 &  9.294 &  7.728 & 28.759 & -67.8 &  25.1 & \textit{-72.2} & 9.78 & 9.19 & 9.78 & 334 \\
				BA$^{12}$  & 1.110 & -7.720 & 11.190 & 40.028 &  59.1 &   1.0 & -69.3 & 7.80 & 19.53 & \textit{12.70 }& 330 \\
				\hhline{=|=|=|=|=|=|=|=|=|=|=|=}
			\end{tabular}
		}	
		\caption{Selection of results. Optimization method used, nodal coordinates, objective function (obj.), stresses in the rods, lengths of the rods, number of evaluations (eval.).$^1$overall minimum objective, $^2$second minimum objective, $^3$best SSA, $^4$best PSO, $^5$best BA, $^6$best solution for $y_A, y_B> 0$, $^7$best ABC $^{8}$worst PSO, $^{9}$worst ABC, violated stress constraint, $^{10}$worst SSA, violated stress constraint, $^{11}$worst NM, violated stress constraint, $^{12}$worst BA, violated length constraint. Violated constraints are written in italics.}
		\label{tab_Ex1_optim}
	\end{center}
\end{table}

\begin{figure}
	\centering
	\includegraphics[scale=1.0]{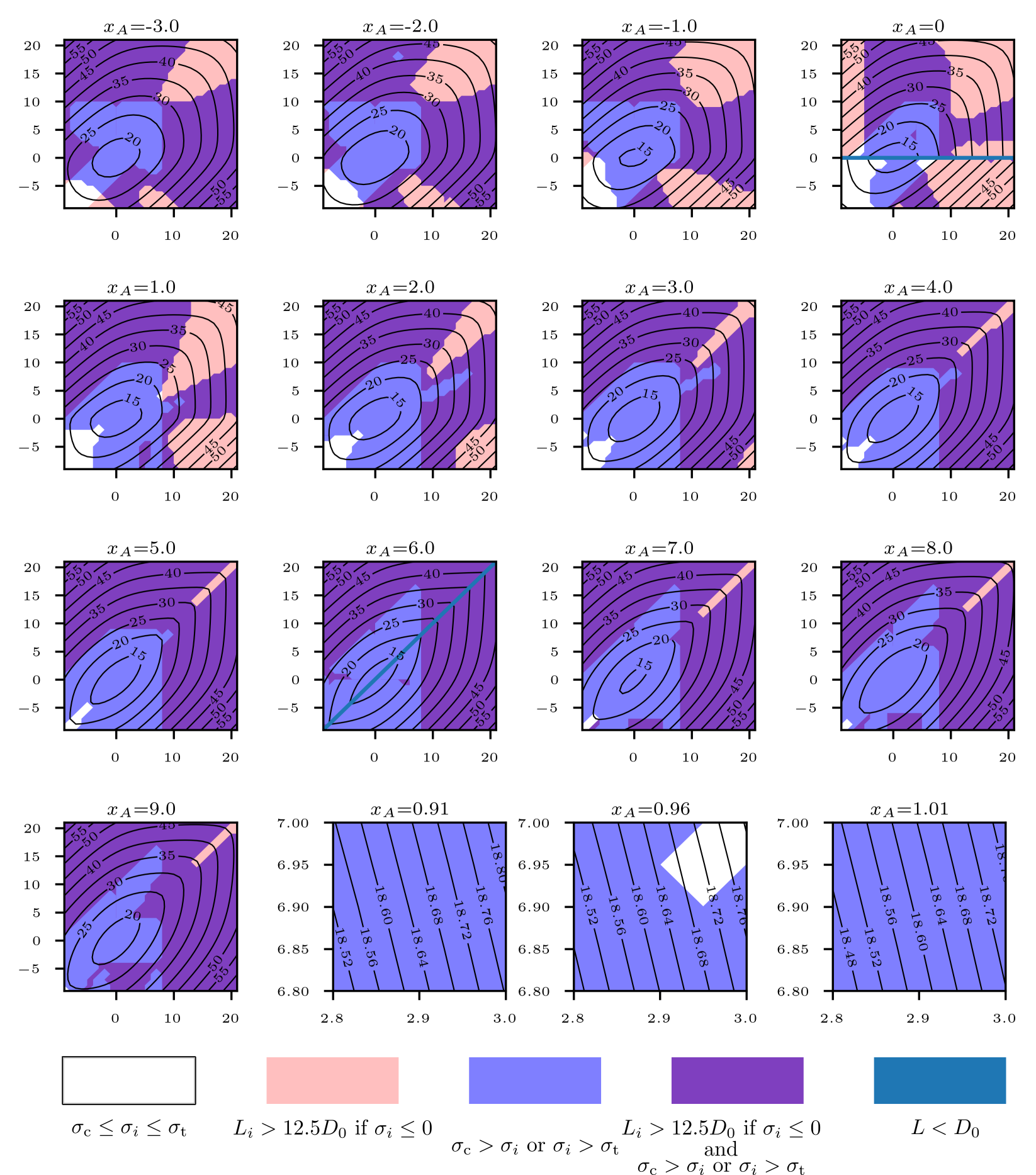}
	\caption{Two-dimensional representation of the design space for different values of $x_A$; white areas indicate feasible configurations, colored areas indicate infeasible ones. Abscissa - $y_A$, ordinate - $y_B$. The contours show the value of the objective. For $x_A$ that are not shown, there are no feasible solutions.} 
	\label{fig_ex1_convergence}
\end{figure}

To get a deeper insight into the problem, the static analysis was performed for all possible combinations of variables in $x_i \in \left[ -15,15 \right]$, $i=1,2,3$, with a step of 1. $x_A$ for all feasible solutions and one infeasible ($x_A=6$) are shown in Fig.~\ref{fig_ex1_convergence}. Due to a rather coarse step used to obtain these results, the constraint Eq.~(\ref{eq_Ex1_opt_prob}c) regarding the minimum length of a rod is represented as a line instead of a narrow area for $x_A=0$ and $x_A=6$. It is interesting to note that this method did not find a single point that fulfills the conditions with $y_A, y_B> 0$ (i.e. in the area of the local minimum). For this reason, a smaller part of variables space around the local minimum ($x_A=0.96$, step size 0.05) is examined. Only two feasible local configurations were then found, which are shown as a white area at $x_A=0.96$. This implies that there are not many such configurations.

\section{Addendum to Example 3}
\label{sec:app3}

\begin{figure}
	\centering
	\includegraphics[scale=0.95]{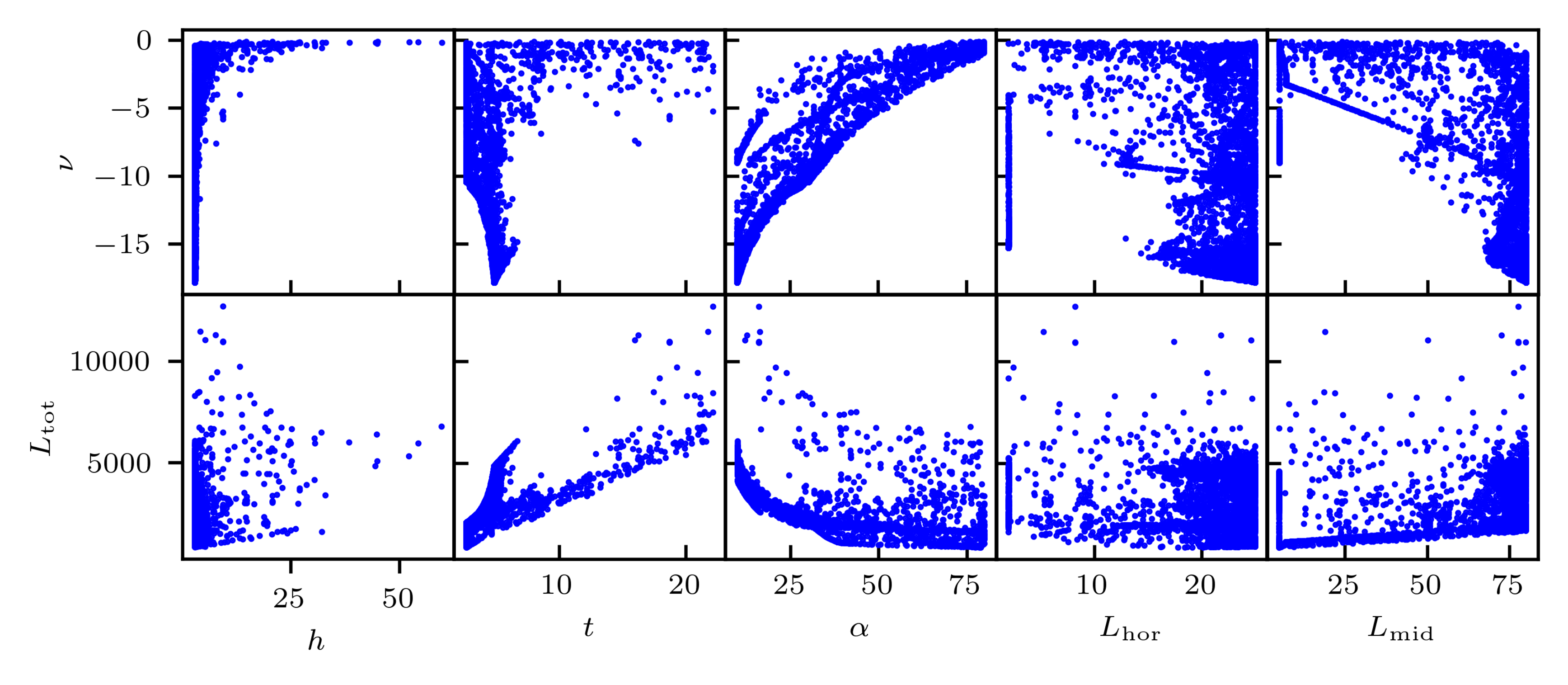}
	\caption{The dependence of objectives on variables. Each dot represents a feasible configuration that the optimization method has found.} 
	\label{fig_ex3_4_correlation}
\end{figure}

The dependence of the objectives on the variables for all feasible solutions is manifested by Spearman's rank correlation coefficients in Tab.~\ref{tab_Ex3_corr} and presented graphically in Fig.~\ref{fig_ex3_4_correlation}. For the Poisson's ratio, the strongest and quadratic positive correlation is visible between $\alpha$ and $\nu$. The second strongest, but negative correlation is exhibited by $t$. The remaining variables do not show such clear trends, but it appears that they also show a linear correlation, although the only positive correlation $\nu$ is obtained for small values of $h$. For the total length of the structure, $\alpha$ again shows the strongest influence, albeit with a negative correlation. The thickness $t$ is also very important. The other variables have a much smaller influence on the change in $L_\mathrm{tot}$ due to smaller correlation coefficients. It is emphasized that this discussion is in line with the previous discussion on the obtained values of the optimal variables.

\begin{table}
	\begin{center}
		\addtolength{\leftskip} {-2cm}
		\addtolength{\rightskip}{-2cm}
		{		\footnotesize
			\begin{tabular}{c|c|c|c|c|c}
				\hhline    {=|=|=|=|=|=|}
				& $h$ & $t$ & $\alpha$ & $L_\mathrm{hor}$ & $L_\mathrm{mid}$ \\
				\hline
				$\nu$ &  0.363 & -0.528 &  0.875 & -0.440 & -0.350 \\
				$L_\mathrm{tot}$ & -0.173 &  0.791 &  -0.926 &  0.254 &  0.147 \\ 
				\hhline{=|=|=|=|=|=|}
			\end{tabular}
		}	
		\caption{Spearman's rank correlation coefficients.}
		\label{tab_Ex3_corr}
	\end{center}
\end{table}

\bigskip

\noindent
\printcredits

\bibliographystyle{cas-model2-names}

\bibliography{SmallSizeParameter}

\end{document}